\documentclass[traditabstract]{aa}
\usepackage{graphicx}
\usepackage{amsmath}
\usepackage{amssymb}

\def\deg{^{\circ}}
\def\df{\textrm{d}}
\newcommand{\ve}[1]{{\rm\bf {#1}}}

\begin{document}

\title{Inferring the magnetic field vector in the quiet Sun}
\subtitle{II. Interpreting results from the inversion of Stokes profiles}

\author{J.M.~Borrero\inst{1} \and P. Kobel\inst{2}}
\institute{Kiepenheuer-Institut f\"ur Sonnenphysik, Sch\"oneckstr. 6, D-79110, Freiburg, Germany. \email{borrero@kis.uni-freiburg.de}
 \and Max-Planck Institut f\"ur Sonnensystemforschung, Max-Planck Str. 2, Katlenburg-Lindau, 37191, Germany. \email{kobel@mps.mpg.de}}
\date{Recieved / Accepted}

\begin{abstract}
{In a previous paper, we argued that the inversion of Stokes profiles applied to spectropolarimetric observations of the solar 
internetwork yield unrealistically large values of the inclination of the magnetic field vector ($\gamma$). This is because
photon noise in Stokes $Q$ and $U$ are interpreted by the inversion code as valid signals, that leads to an overestimation of the
transverse component $B_\perp$, thus the inclination $\gamma$. However, our study was based
on the analysis of linear polarization signals that featured only uncorrelated noise. In this paper, we develop this idea further
and study this effect in Stokes $Q$ and $U$ profiles that also show correlated noise. In addition, we extend our study to the 
three components of the magnetic field vector, as well as the magnetic filling factor $\alpha$. With this, we confirm the 
tendency to overestimate $\gamma$ when inverting linear polarization profiles that, although non-zero, are still below the noise level.
We also establish that the overestimation occurs mainly for magnetic fields that are nearly vertical $\gamma \lesssim 20\deg$. This indicates
that a reliable inference of the inclination of the magnetic field vector cannot be achieved by analyzing only Stokes $I$ and $V$. In addition, 
when inverting Stokes $Q$ and $U$ profiles below the noise, the inversion code retrieves a randomly uniform distribution of the azimuth 
of the magnetic field vector $\phi$. To avoid these problems, we propose only inverting Stokes profiles for which the linear polarization
signals are sufficiently above the noise level. However, this approach is also biased because, in spite of allowing for a very accurate retrieval
of the magnetic field vector from the selected Stokes profiles, it selects only profiles arising from highly inclined magnetic fields.}
\end{abstract}

\maketitle

\section{Introduction}
\label{section:intro}

Spectropolarimetry, which is the study of the polarization properties of the light observed in spectral lines, is the most developed and 
widely used tool for retrieving the magnetic properties of the solar plasma and other astrophysical objects (Stenflo 2002; 
Mathys 2002 and references therein). The inference of the magnetic field vector from spectropolarimetric observations
is performed through the inversion of the radiative transfer equation for polarized light (del Toro Iniesta 2003a; Bellot Rubio 2004;
Ruiz Cobo 2007). However, as the module of the magnetic field vector decreases, the observed signals disappear below
the level of the photon noise, making the inference of the magnetic field very difficult and plagued with problems and uncertainties.
This is, for instance, the case for the areas on the solar surface referred to as the \emph{solar internetwork}. Here, the polarization signals are 
so weak that for a long time the internetwork was thought to be void of magnetic fields.\\

With the advancements in the sensitivity of the polarimeters and the higher spatial resolution of the observations
achievable with adaptive-optic systems and large-aperture telescopes, polarization signals are now routinely detected everywhere 
in the internetwork (Lites et al. 1996). This demonstrates that these regions are truly pervaded by magnetic fields. Unfortunately, 
the signals are barely above the noise level, thus it has not yet been possible to fully characterize the magnetic field vector 
in these regions. This has led to a long-standing controversy about the distribution of the module of the magnetic field vector
in the internetwork (Dom{\'\i}nguez Cerde\~na et al. 2003; Socas-Navarro \& Lites 2004; S\'anchez Almeida 2005; Mart{\'\i}nez Gonz\'alez et al. 2006;
Asensio Ramos et al. 2007; L\'opez Ariste et al. 2007; Socas-Navarro et al. 2008; see also references therein).\\

More recently, further discrepancies have emerged about the angular distribution of the magnetic field vector 
(Lites et al. 2007, 2008; Orozco Su\'arez et al. 2007a, 2007b; Mart{\'\i}nez Gonz\'alez et al. 2008; Asensio Ramos 
2009; Stenflo 2010; Ishikawa \& Tsuneta 2011; Borrero \& Kobel 2011). In Borrero \& Kobel (2011; hereafter referred to 
as paper I), we argued that the photon noise leads to a systematic
overestimation of the inclination of the magnetic field vector. We reached this conclusion because when employing
only vertical magnetic fields to synthesize Stokes profiles, these could be retrieved as highly inclined ones due to the sole effect
of the photon noise. Since those tests were carried out with vertical magnetic fields, the linear polarization signals (Stokes $Q$
and $U$) were zero. This limits the validity of the tests, as the inversion code analyzes only linear polarization profiles
featuring uncorrelated noise. It is therefore worth considering the case where small linear-polarization signals (due to non-vertical
magnetic fields) can be hidden below the noise. This introduces correlations that could be employed by the inversion algorithms to
retrieve useful information about the inclination of the magnetic field vector. This paper is devoted to addressing this particular situation.
To this end, we employ uniform distributions of the magnetic field vector (Section~\ref{section:uniformpdf}) to produce a large database
of Stokes profiles. Therefore, unlike paper I, the resulting Stokes profiles will 
have non-zero linear polarization profiles. After adding noise to these profiles, we apply an inversion algorithm 
(Section~\ref{section:inversions}) that attempts to retrieve the original distribution of the magnetic field vector. Our results are
described in Sections~\ref{section:discussion} and \ref{section:correlations}. Section~\ref{section:conclu} summarizes our findings.\\

\section{Synthesis with uniform distributions}
\label{section:uniformpdf}

To study the effect of the photon noise and selection criterion, we performed several numerical 
experiments employing a probability distribution function that is parametrized by the physical 
parameters of a Milne-Eddington atmosphere (ME). These physical parameters are denoted as $\ve{X}$

\begin{equation}
\ve{X} = [\ve{B}, \alpha, V_{\rm los}, \ve{T}] \;,
\label{equation:x}
\end{equation}

\noindent where $\ve{B}$ refers to the magnetic field vector, $\alpha$ is the so-called magnetic
filling factor and denotes the fractional area within the pixel that is occupied by magnetized plasma, $V_{\rm los}$
is the line-of-sight component of the velocity vector, and, finally, the vector $\ve{T}$ considers all thermodynamic
quantities (source function, Doppler width, and so forth). More details about these parameters and the Milne-Eddington 
approximation can be found in Landi Degl'Innocenti (1992) and del Toro Iniesta (2003b). Here we write the
probability distribution function for the M-E parameters $\ve{X}$ in the form\\ 

\begin{equation}
\mathcal{P}(\ve{X})\df\ve{X} = \mathcal{P}_1(\ve{B},\alpha) \mathcal{P}_2(\ve{T},V_{\rm los}) \df\ve{B}\df\alpha
\df\ve{T} \df V_{\rm los} \;,
\label{equation:pdf}
\end{equation}

\noindent where the thermodynamic and kinematic parameters, $\ve{T}$ and $V_{\rm los}$, are assumed to be statistically
independent of the magnetic ones $\ve{B}$ and $\alpha$. The distribution function of the former parameters
$\mathcal{P}_2(\ve{T},V_{\rm los})$ is the one obtained from the inversion of map A in Borrero \& Kobel (2011), 
so that we employ values that are representative of the quiet Sun. The term that contains the properties of the magnetic field, 
$\mathcal{P}_1(\ve{B},\alpha)$, has been modeled as a slightly-modified uniform distribution, expressed as\\

\begin{eqnarray}
\mathcal{P}_1(\ve{B},\alpha) \df\ve{B} \df\alpha = \frac{1}{2 \pi^2 B_0}\textrm{H}(B-B_0) \df B \df\gamma \df\phi 
\df\alpha \;,
\label{equation:pdfuniform}
\end{eqnarray}

\noindent which gives the probability of finding a magnetic field vector \ve{B} whose module is between $B$ and $B+\df B$, 
whose inclination (with respect to the observer's line-of-sight) is between $\gamma$ and $\gamma+\df\gamma$, and whose azimuth 
(in the plane perpendicular to the observer's line-of-sight) is between $\phi$ and $\phi+\df\phi$. On the right-hand-side 
of this equation, the term $\textrm{H}(B-B_0)$ also refers to the complementary Heaviside function\\

\begin{eqnarray}
\textrm{H}(B-B_0) & = \begin{cases} 1, & \textrm{if}\; B < B_0 \\ 0, & \textrm{if}\; B > B_0 \end{cases} \;,
\label{equation:heaviside}
\end{eqnarray} 

\noindent where $B_0$ is taken as $B_0=200$ G in order to make our experiment representative of weak
field regions in the solar surface. In addition to the value of $B_0$, we do not make any further attempts to
employ a more realistic model for the solar internetwork since our aim is not to investigate particular distribution functions, but rather
to study the effect of the inversion, photon noise and selection criteria in the inference of the magnetic field vector in a general
way.\\

Once we have the probability distribution function given by Eqs.~\ref{equation:pdf}-\ref{equation:pdfuniform}-\ref{equation:heaviside}, we solve
the radiative transfer equation in order to create a large database ($=2\times 10^6$) of synthetic/theoretical
Stokes profiles of the \ion{Fe}{I} 6302.5 {\AA} ($g_{\rm eff}=2.5$) spectral line. The effective Land\'e factor 
of the atomic transition calculated under the LS coupling scheme is indicated by $g_{\rm eff}$. The Stokes profiles
are synthesized with the VFISV (Very Fast Inversion of the Stokes Vector) inversion code of Borrero et al. (2010).
To these profiles, photon noise is added as a normally distributed random variable with a standard deviation
of $\sigma = 10^{-3}, 3 \times 10^{-4}$. These two values mimic the noise levels of maps A and C in Borrero \& 
Kobel (2011). Once the noise is added, the resulting Stokes profiles are taken as real observations and
inverted in order to retrieve the original atmospheric parameters $\ve{X}$. This step is described in the next section.\\

\section{Inversions of synthetic Stokes profiles}
\label{section:inversions}

The Stokes profiles for \ion{Fe}{I} 6302.5 {\AA} synthesized in the previous section are now inverted with the
same VFISV code, but now running in inversion mode instead of synthesis mode. At the start,
VFISV solves the radiative transfer equation for polarized light in the Milne-Eddington (ME) approximation using a
set of initial values for the physical parameters: $\ve{X}_0$. The solution of the radiative transfer equation yields 
theoretical or synthetic Stokes profiles that are compared to the ones synthesized in the previous section
(those where photon noise had been added). The initial values of $\ve{X}_0$ are then iteratively modified until the 
best possible fit between the theoretical/synthetic and observed Stokes vector is reached. The final $\ve{X}_{\rm f}$ that 
achieves the best fit is then assumed to be the real one present in the solar atmosphere. In this work we will focus
mainly on the magnetic field vector $\ve{B}$ and magnetic filling factor $\alpha$. For more details about how the inversion 
is performed, additional free parameters of the inversion, and treatment of the filling factor, we refer the reader 
to Borrero et al. (2010) and Borrero \& Kobel (2011; paper I). We note that, since we are using the same type of atmospheric
model in the synthesis and inversion (ME atmospheres), the experiments carried out in this paper do not address the 
systematic errors introduced by the choice of the wrong atmospheric model, such as those introduced when inverting asymmetric 
Stokes profiles using a ME atmospheric model\footnote{Other sources of systematic errors that are not being considered, as 
they are beyond the scope of our study, are the effects of the spatial resolution, spectral resolution, etcetera. For instance,
Borrero et al. (2007) carried out such a study for the particular case of the Helioseismic and Magnetic Imager (HMI)
instrument.}.\\

Once $B$, $\gamma$, $\phi$ and $\alpha$ had been retrieved from each Stokes profiles of the original database, we applied 
two different selection criteria. In the first criteria, we select for the analysis profiles in the database
where the maximum of the absolute value (for all wavelengths) in \emph{any} of the polarization signals (Stokes $Q$, $U$, or $V$) is larger than 4.5 times
the noise level max$|Q(\lambda), U(\lambda), V(\lambda)| \ge 4.5 \sigma$. This is equivalent to selecting pixels where the signal-to-noise
ratio ($S/R$) in the polarization profiles is 4.5 or better, i.e., $S/R \ge 4.5$. Hereafter we refer to this criteria as
$S/R_{\rm quv}$. In the second criteria we select those profiles within the database where the maximum of the absolute
value (for all wavelengths) in the linear polarization signals (Stokes $Q$, $U$) is larger than 4.5 times the noise level
max$|Q(\lambda), U(\lambda)| \ge 4.5 \sigma$. This criteria will  be referred to as $S/R_{\rm qu}$.\\

The reason behind the choice of these two different selection criteria is the following. The first criterion 
(selection of profiles where $S/R_{\rm quv} \ge 4.5$) was adopted so we can draw parallelisms with both the results
presented in paper I and Orozco Su\'arez et al. (2007a, 2007b). Owing to the intrinsic amplitude of 
Stokes $Q$, $U$, and $V$, this criterion selects mostly pixels where only the circular 
polarization $V$ possesses a $S/R \ge 4.5$. The actual percentages depend on the distribution of the
magnetic field vector and the level of noise. For instance, taking a uniform distribution of $B$, $\gamma$,
and $\phi$ (Equation~\ref{equation:pdfuniform}) and considering a noise-level of $\sigma = 10^{-3}$,
only 21.2 \% of the profiles selected with the $S/R_{\rm quv}$-criterion possess linear polarization signals
(Stokes $Q$ and $U$) with peak-values above $4.5\sigma$. In the remaining 78.8 \% of the profiles, only
Stokes $V$ is above $4.5\sigma$. The 21.2 \% increases up to 39.1 \% when considering a noise level of
$\sigma=3\times 10^{-4}$. In paper I, the maps that had equivalent levels of noise featured different
percentages of profiles: in the map with $\sigma=10^{-3}$, only about 8.8 \% of the profiles
selected with the $S/R_{\rm quv}$-criterion had linear polarization profiles above $4.5\sigma$. This number
increased to 38.8 \% in the map with $\sigma=3\times 10^{-4}$. As already demonstrated in that paper (see also
Mart{\'\i}nez-Gonz\'alez et al. 2011), the inversion of profiles selected with the $S/R_{\rm quv}$-criterion yields values 
for the inclination of the magnetic field vector, $\gamma$, that are largely overestimated. This happens 
because most of the Stokes profiles have linear polarization signals that are either below or close to the noise level. This effect 
decreases as the noise in the polarization profiles goes down so that the linear polarization signal rises above 
the noise. However, since both Stokes $Q$ and $U$ remain below $10^{-4}$ for transverse fields up 
to $B_\perp \approx 20-30$ G, the overestimation in the inclination remains significant even at such low levels of noise.\\

In paper I, we proposed, as an alternative solution, selecting for the inversion only pixels where the peak 
in the $Q$ or $U$ signals is at least 4.5 times higher than the noise i.e., where the $S/R$ is higher than 
4.5 in linear polarization. We also followed this approach in applying the second criterion 
(see also Asensio Ramos 2009). In paper I, we anticipated that this approach has the advantage of retrieving more 
trustworthy values of $\gamma$, $B_\perp$, and $B$ but we did not quantify it.\\

\begin{figure*}
\begin{center}
\begin{tabular}{cc}
\includegraphics[width=8cm]{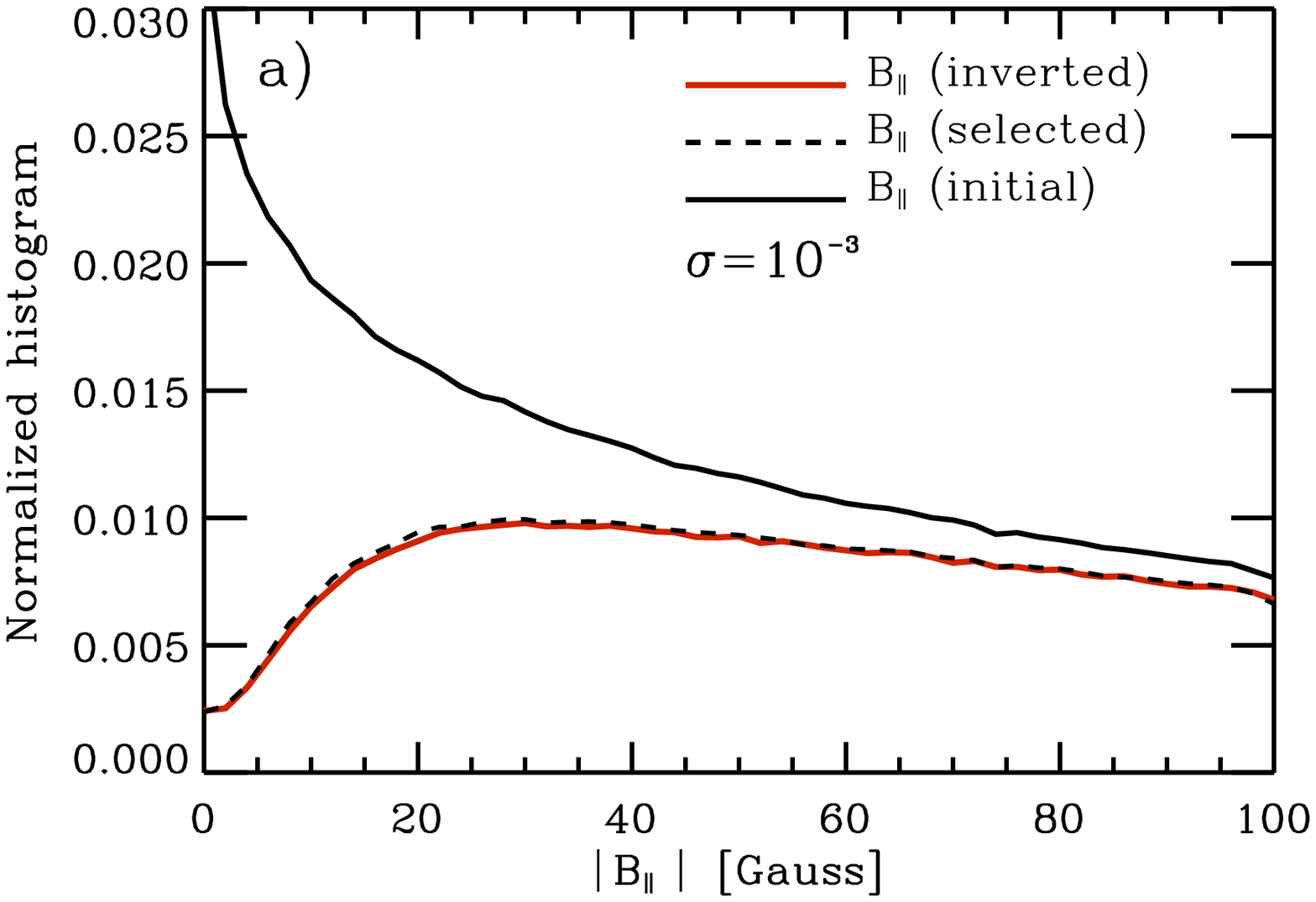} &
\includegraphics[width=8cm]{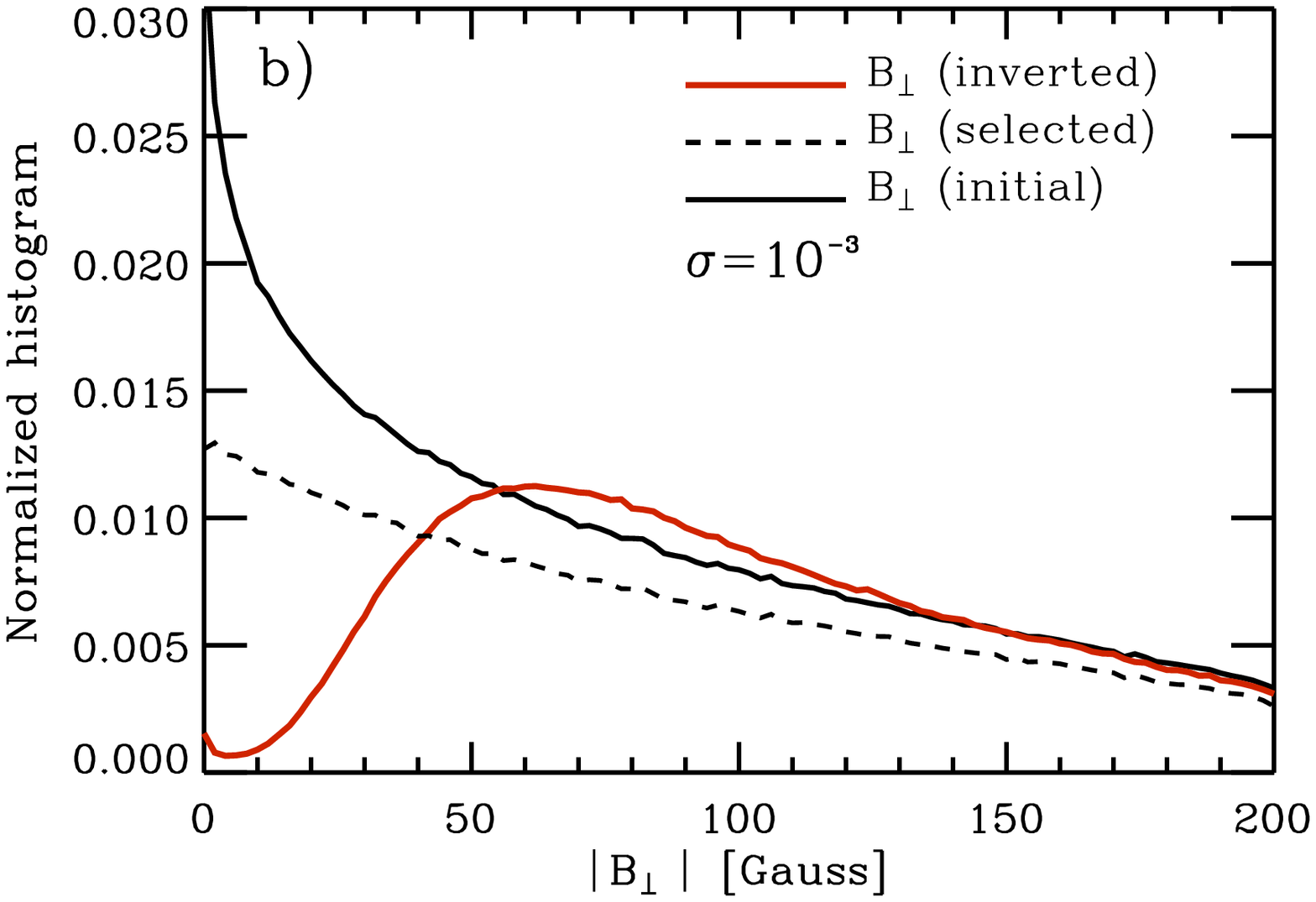} \\
\includegraphics[width=8cm]{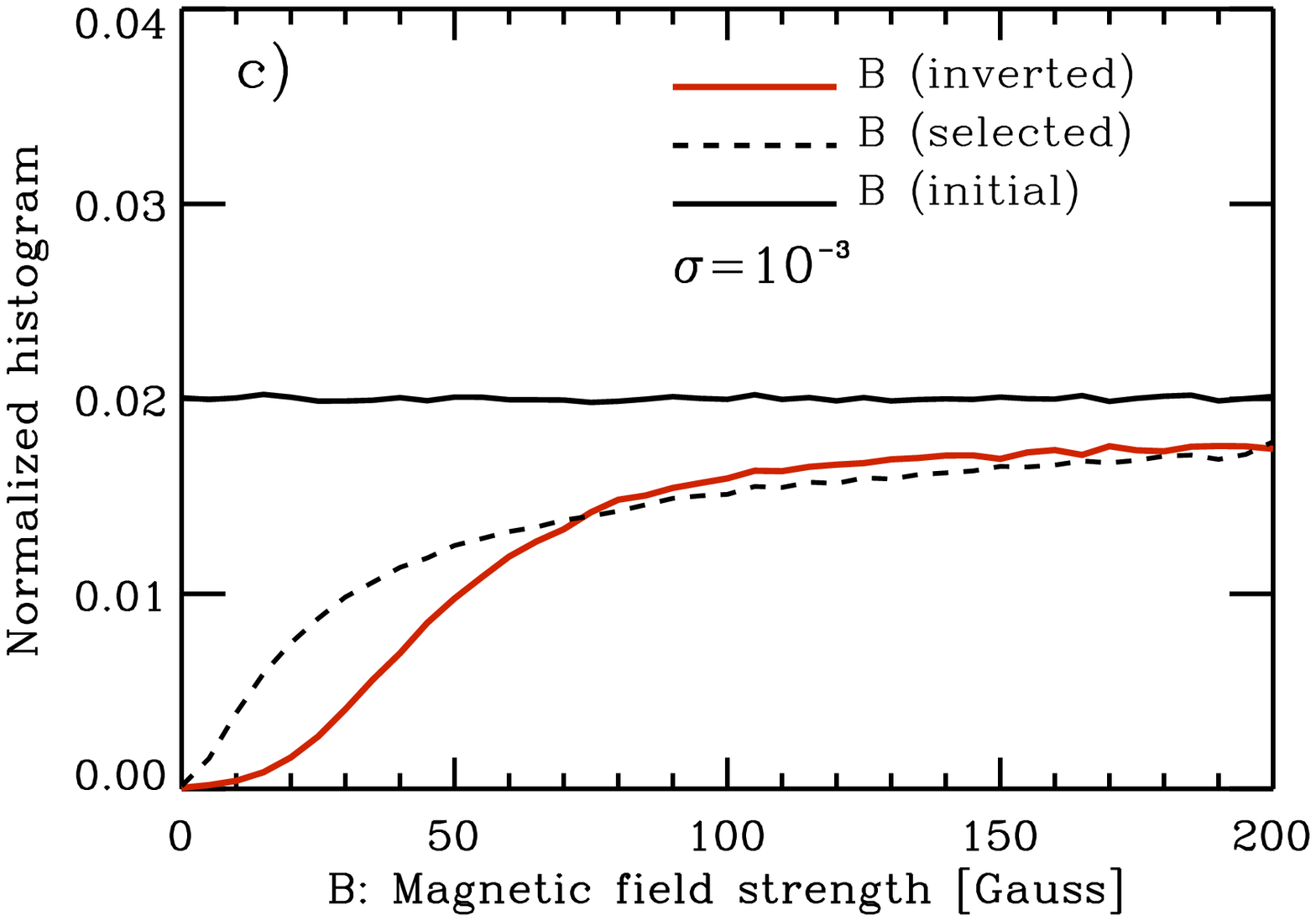} &
\includegraphics[width=8cm]{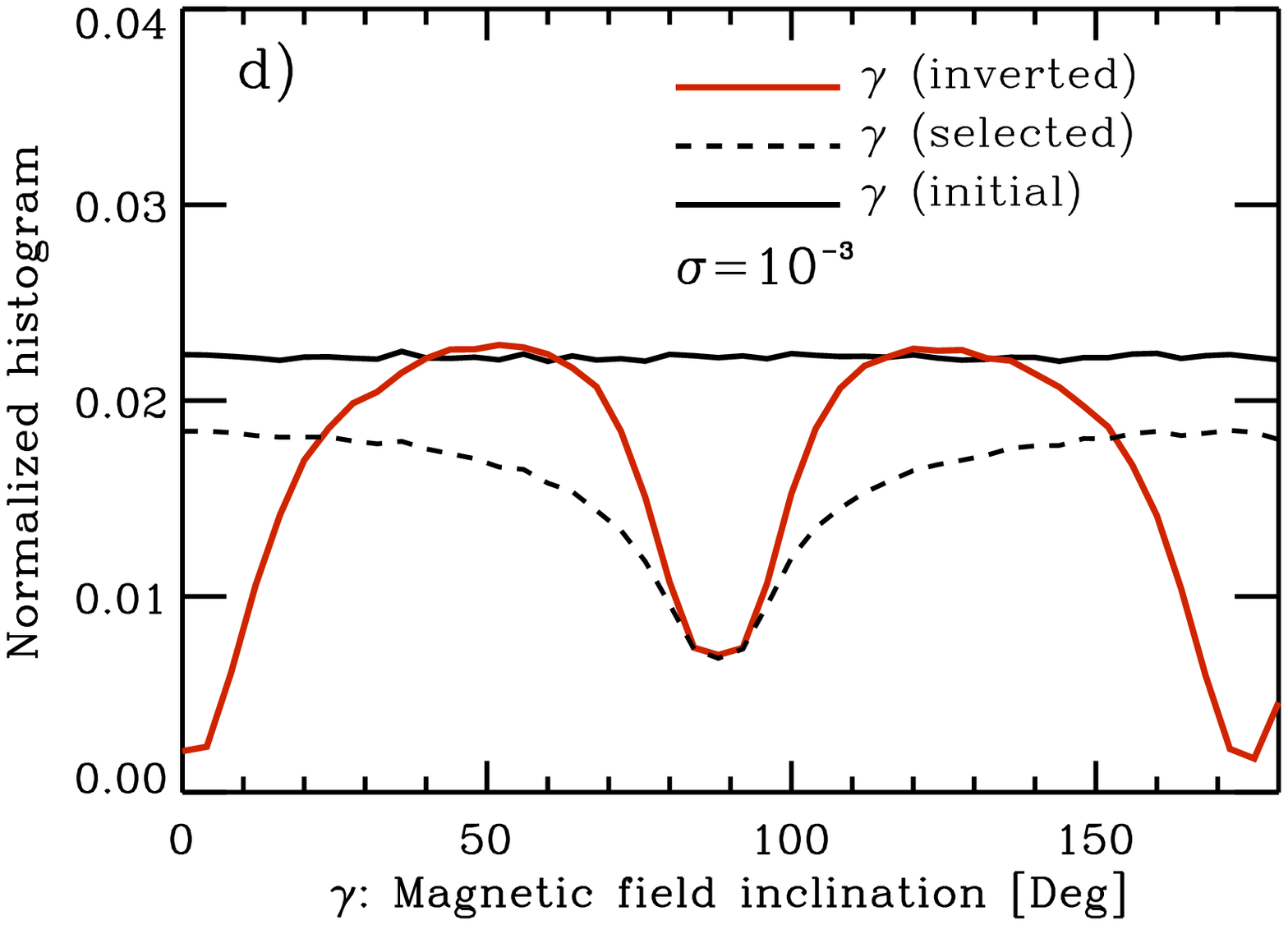} \\
\includegraphics[width=8cm]{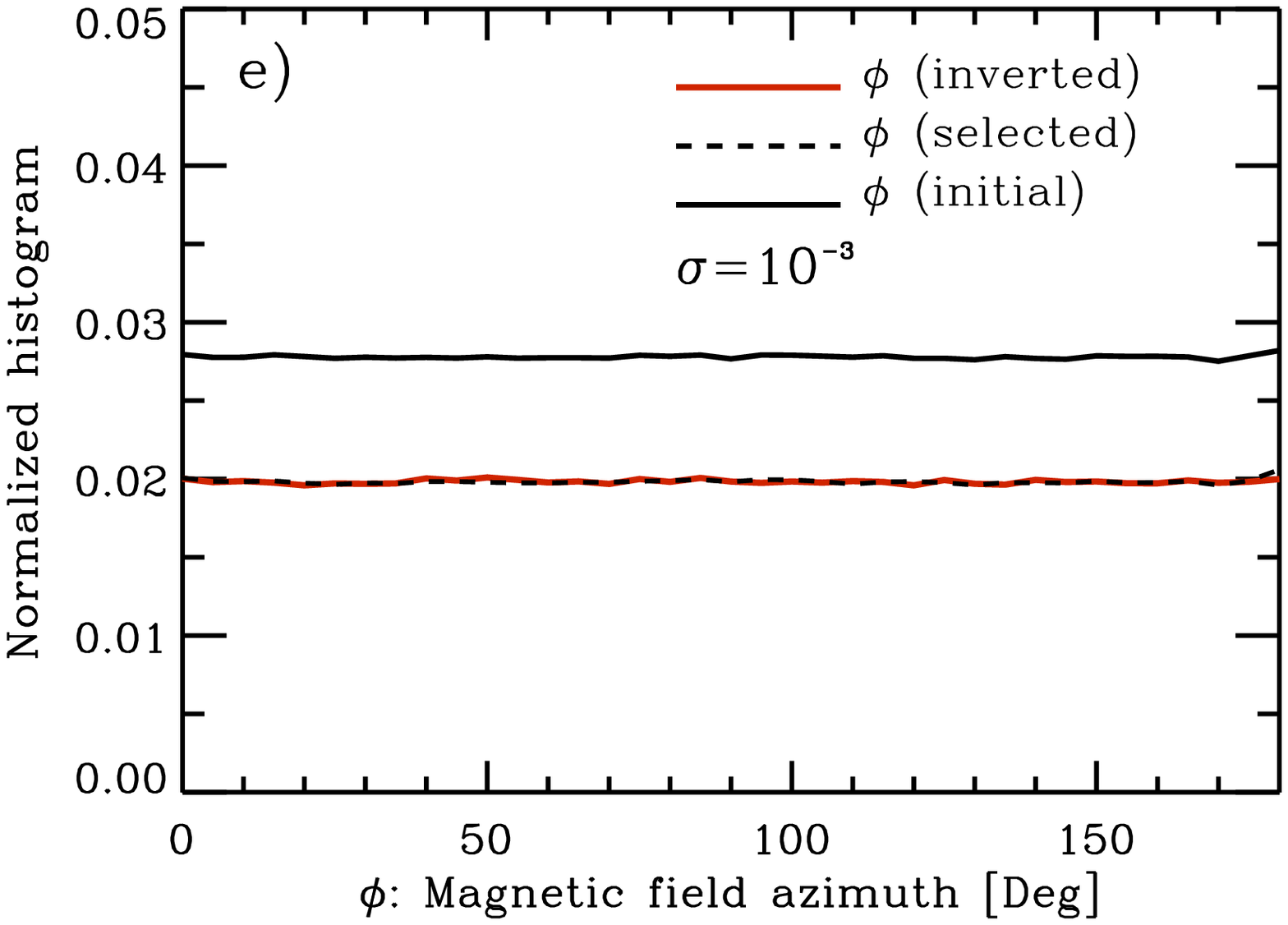} &
\includegraphics[width=8cm]{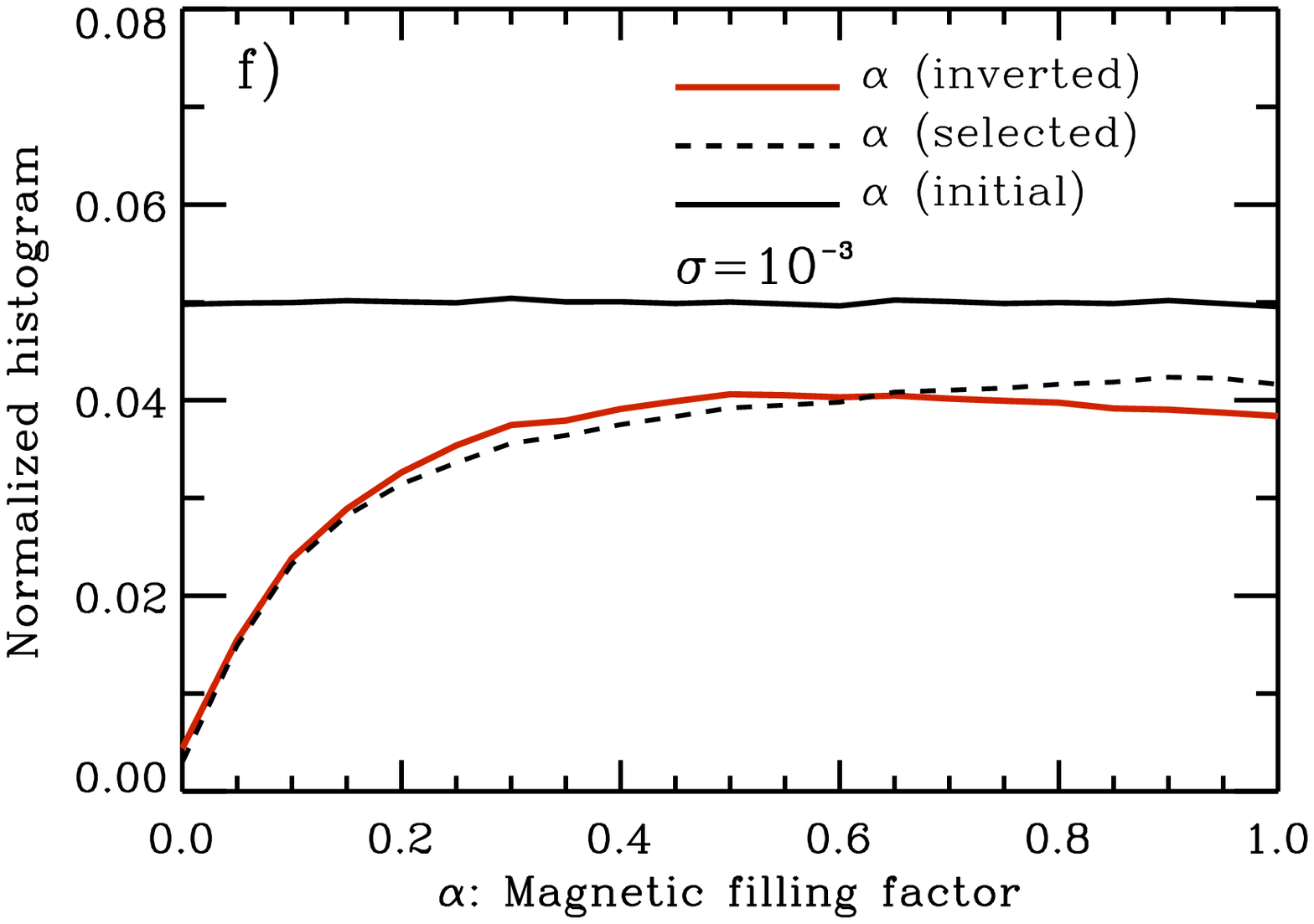} \\
\end{tabular}
\end{center}
\caption{Results of the numerical experiment of the synthesis and inversion of Stokes profiles. The level
noise added is $\sigma=10^{-3}$. Solid-black lines indicate the original distributions employed in the 
synthesis of $2\times 10^{6}$ Stokes profiles (Equation~\ref{equation:pdfuniform}). Dashed lines show
the distributions obtained employing only the Stokes profiles that are selected with the $S/R_{\rm quv}$-criterion.
The red lines display the distributions obtained from the inversion of the selected profiles: {\bf a})
the absolute value of the component of the magnetic field vector that is parallel to the observer's line-of-sight $| B_\parallel | =| B\cos\gamma |$;
{\bf b}) the component of the magnetic field vector that is perpendicular to the observer's line-of-sight 
$B_\perp=B\sin\gamma$; {\bf c}) the magnetic field strength or module of the magnetic field vector $B$; {\bf d}) 
the inclination of the magnetic field vector with respect to the observer's line-of-sight $\gamma$; {\bf e}) 
the azimuthal angle of the magnetic field vector on the plane that is perpendicular to the observer's line-of-sight $\phi$; 
{\bf f}) the magnetic filling factor $\alpha$. In all panels, the dashed-black and solid-red curves are normalized to 
the quotient of the number of profiles selected with each particular criteria to the total number of profiles synthesized 
($2\times 10^6$).}
\label{figure:highquv}
\end{figure*}

\begin{figure*}
\begin{center}
\begin{tabular}{cc}
\includegraphics[width=8cm]{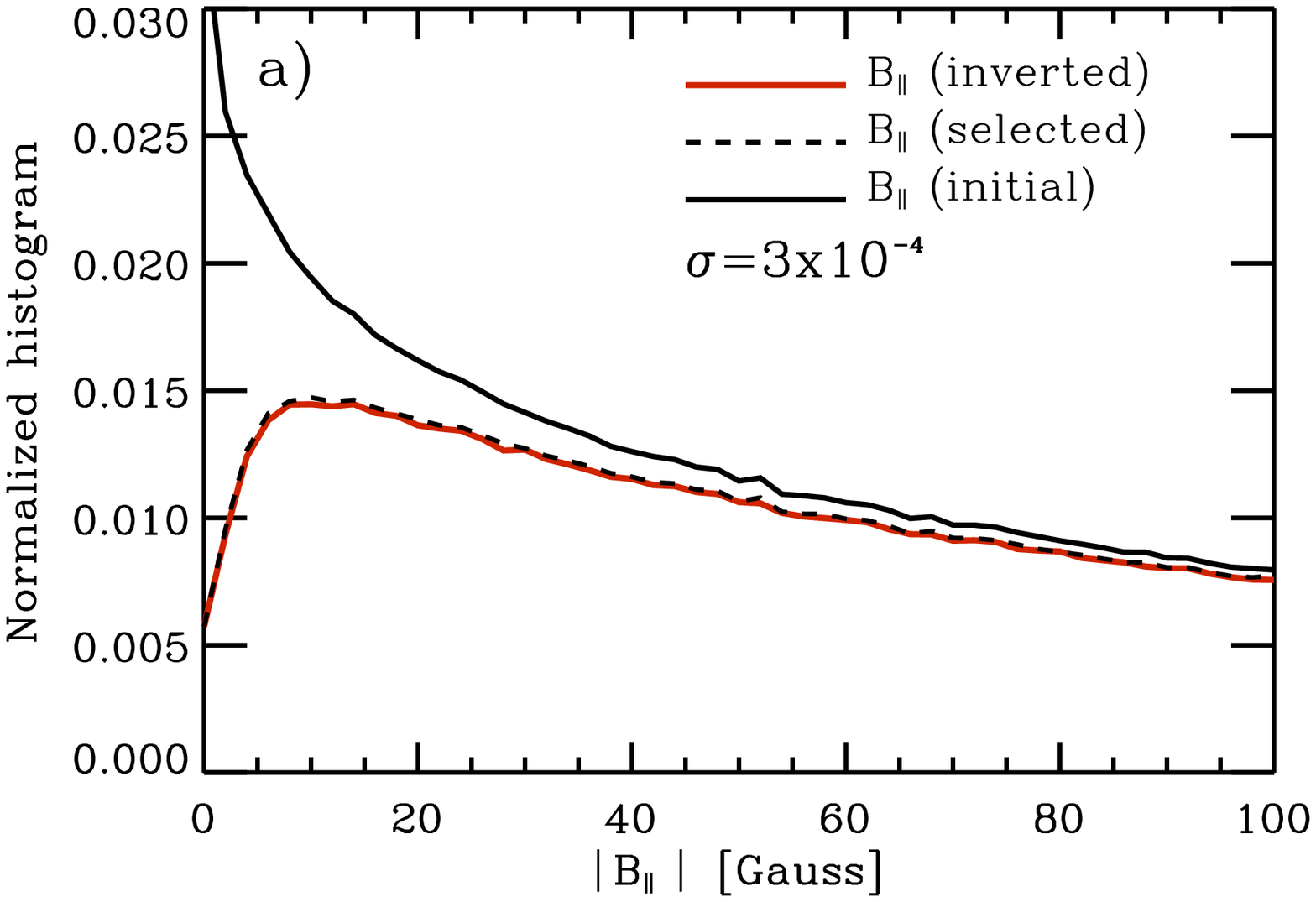} &
\includegraphics[width=8cm]{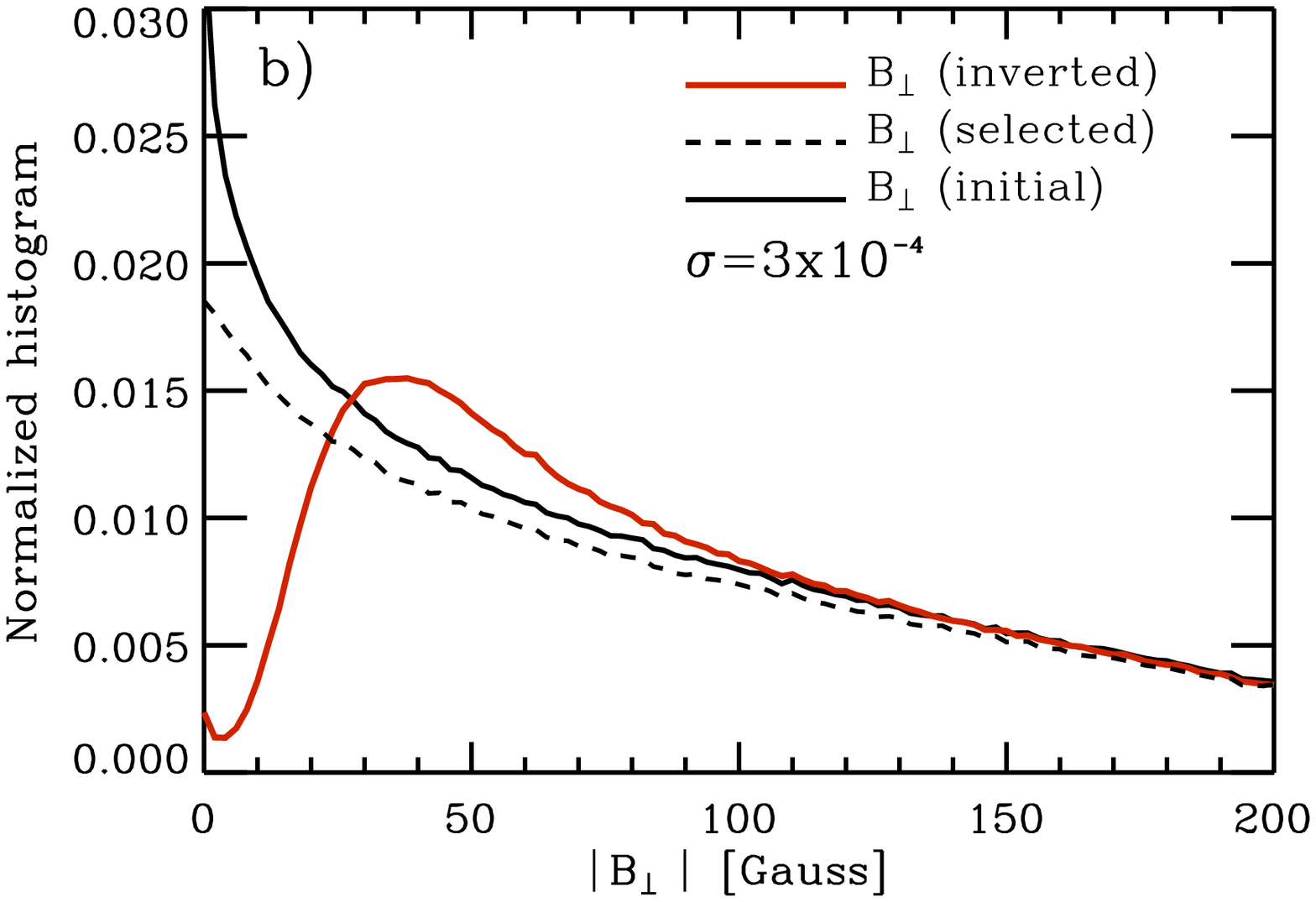} \\
\includegraphics[width=8cm]{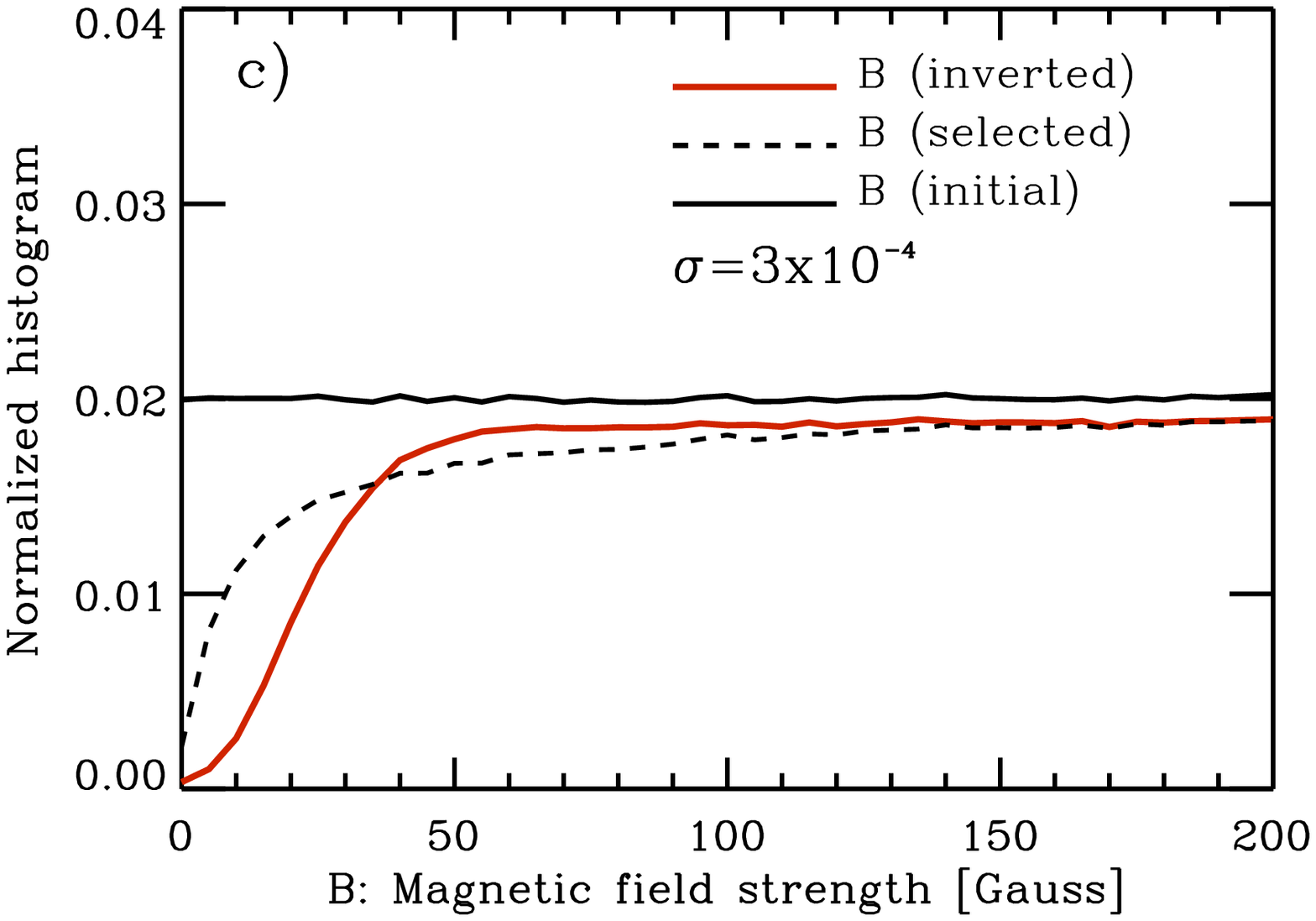} &
\includegraphics[width=8cm]{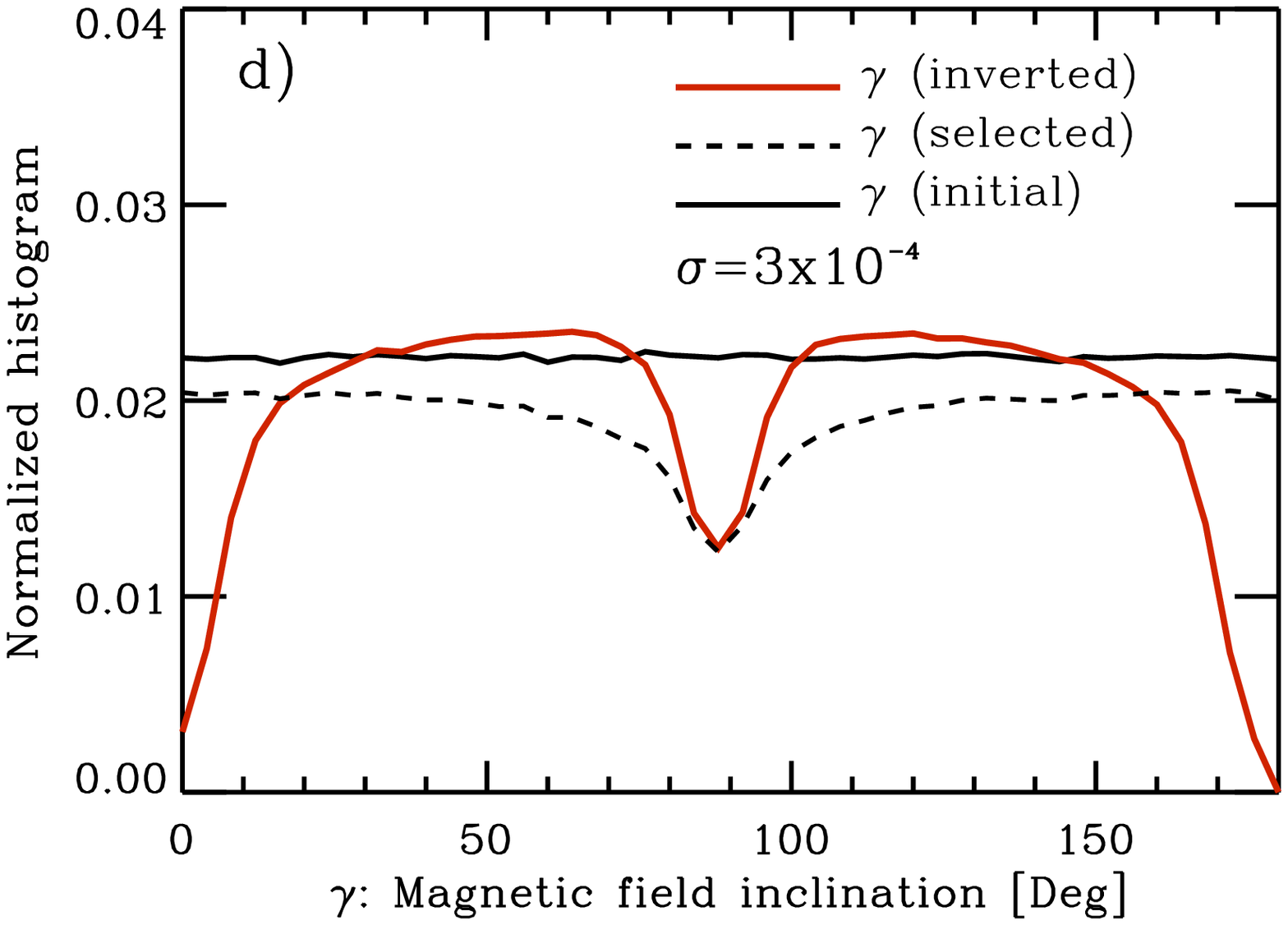} \\
\includegraphics[width=8cm]{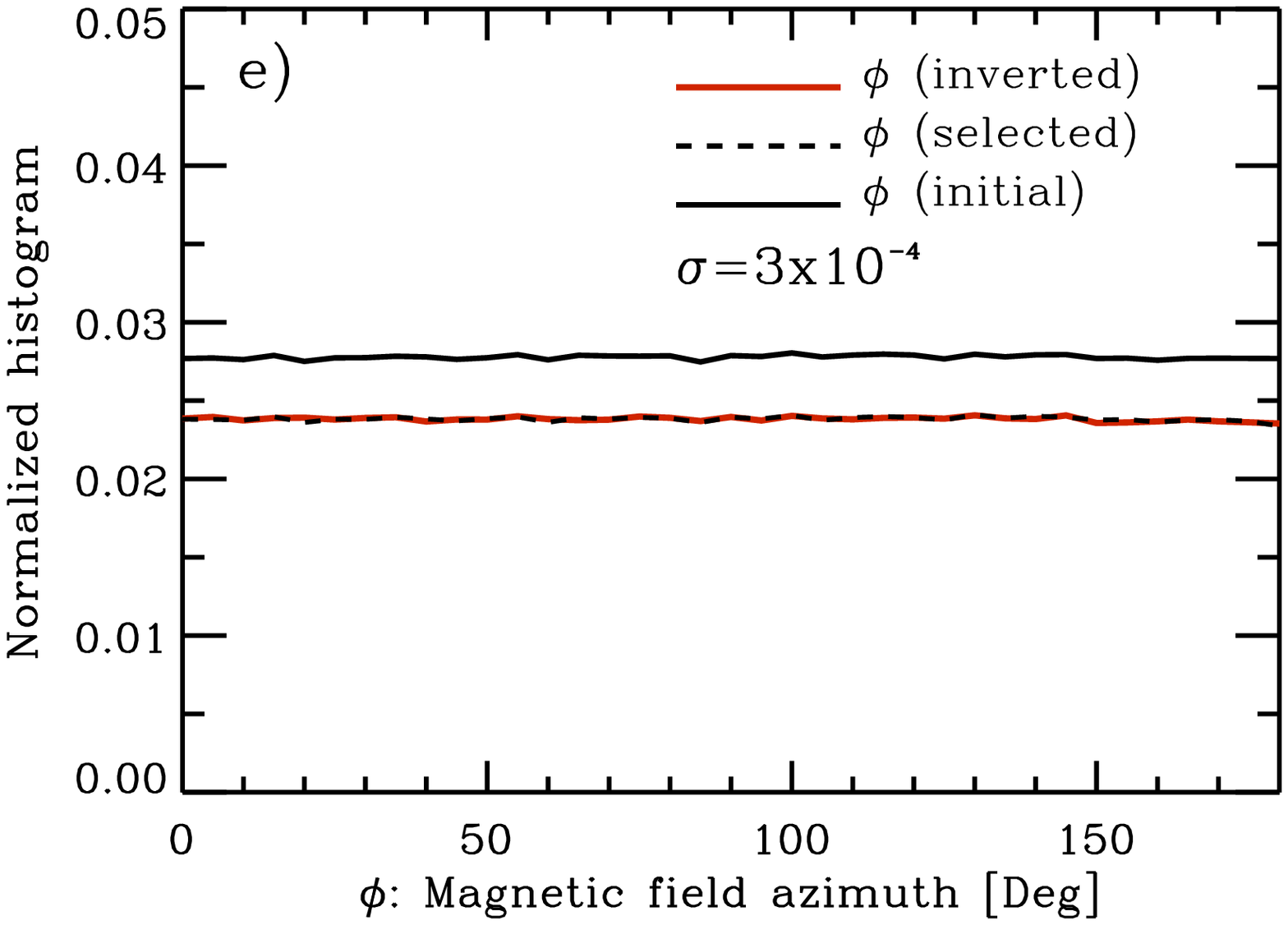} &
\includegraphics[width=8cm]{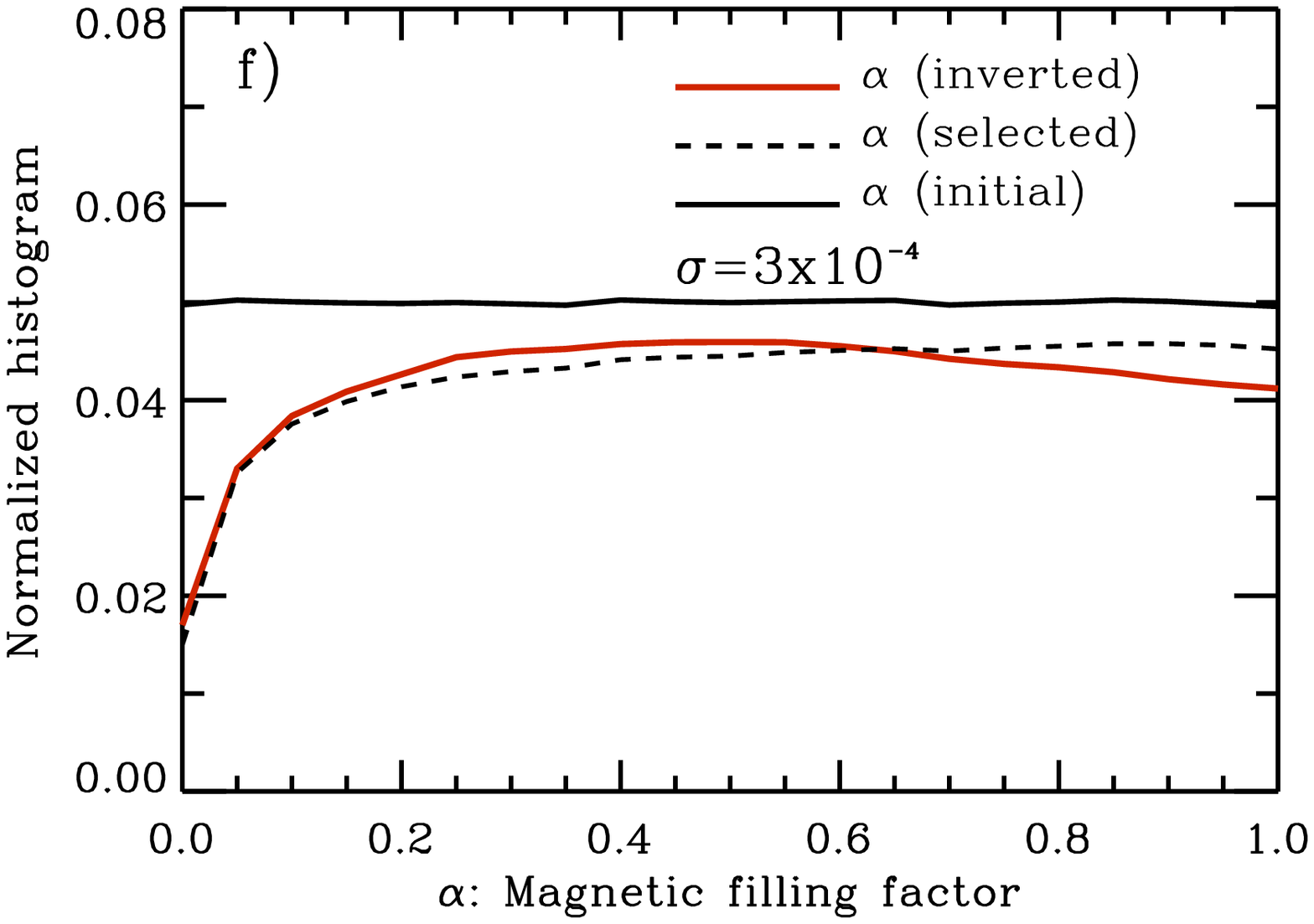} \\
\end{tabular}
\end{center}
\caption{Same as Figure~\ref{figure:highquv} but using a photon noise $\sigma=3\times 10^{-4}$ and the $S/R_{\rm quv}$-criterion.}
\label{figure:lowquv}
\end{figure*}

\begin{figure*}
\begin{center}
\begin{tabular}{cc}
\includegraphics[width=8cm]{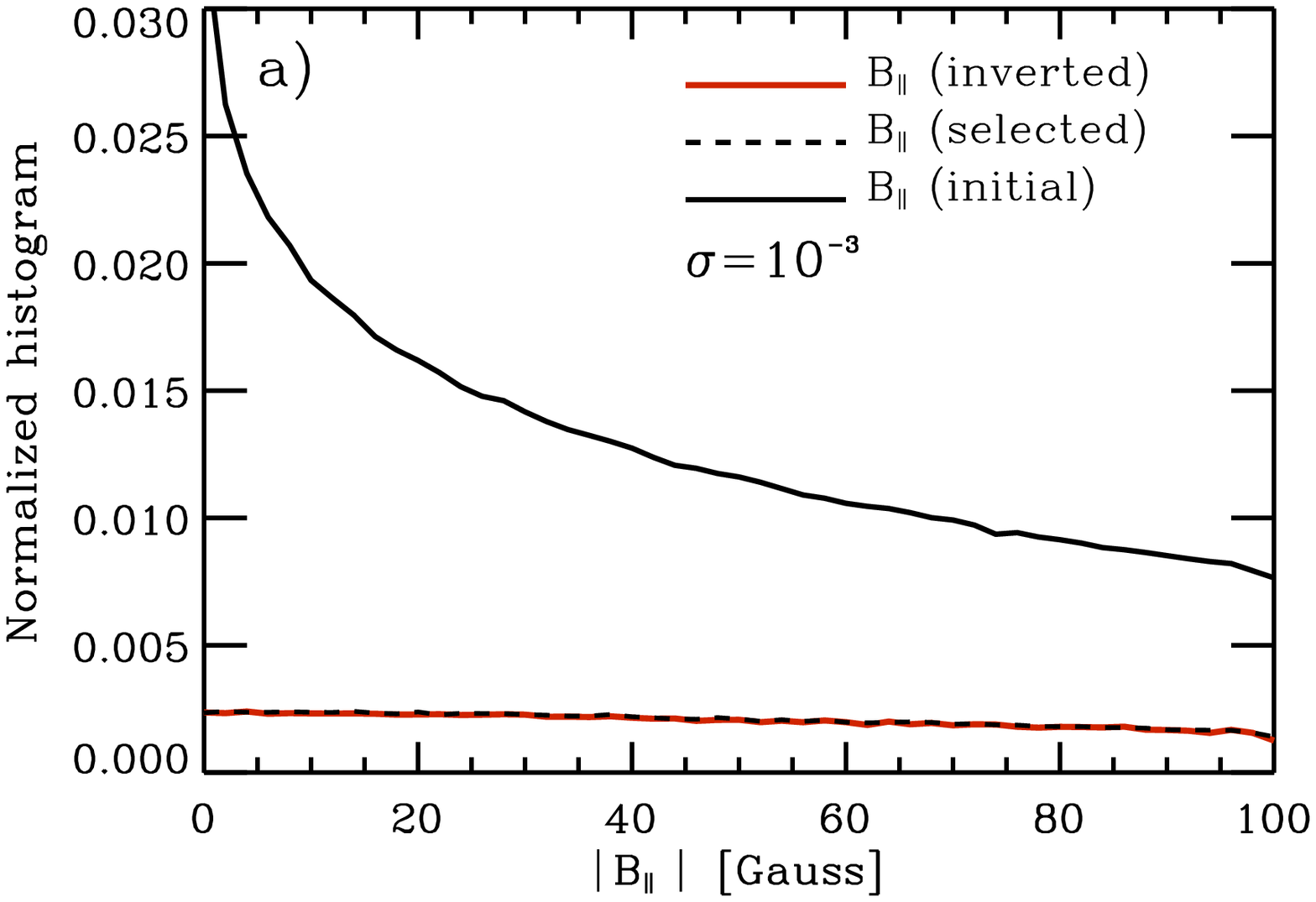} &
\includegraphics[width=8cm]{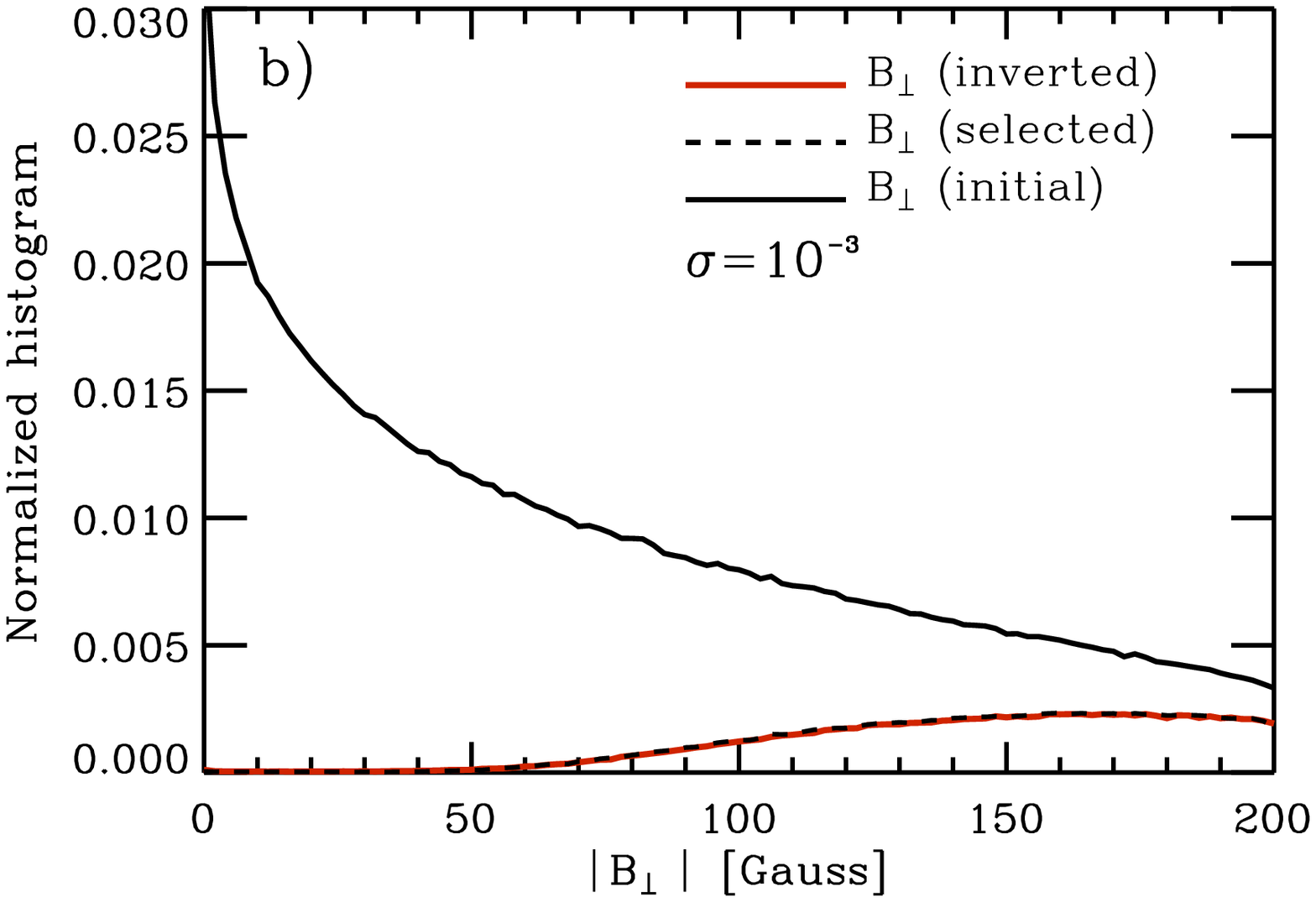} \\
\includegraphics[width=8cm]{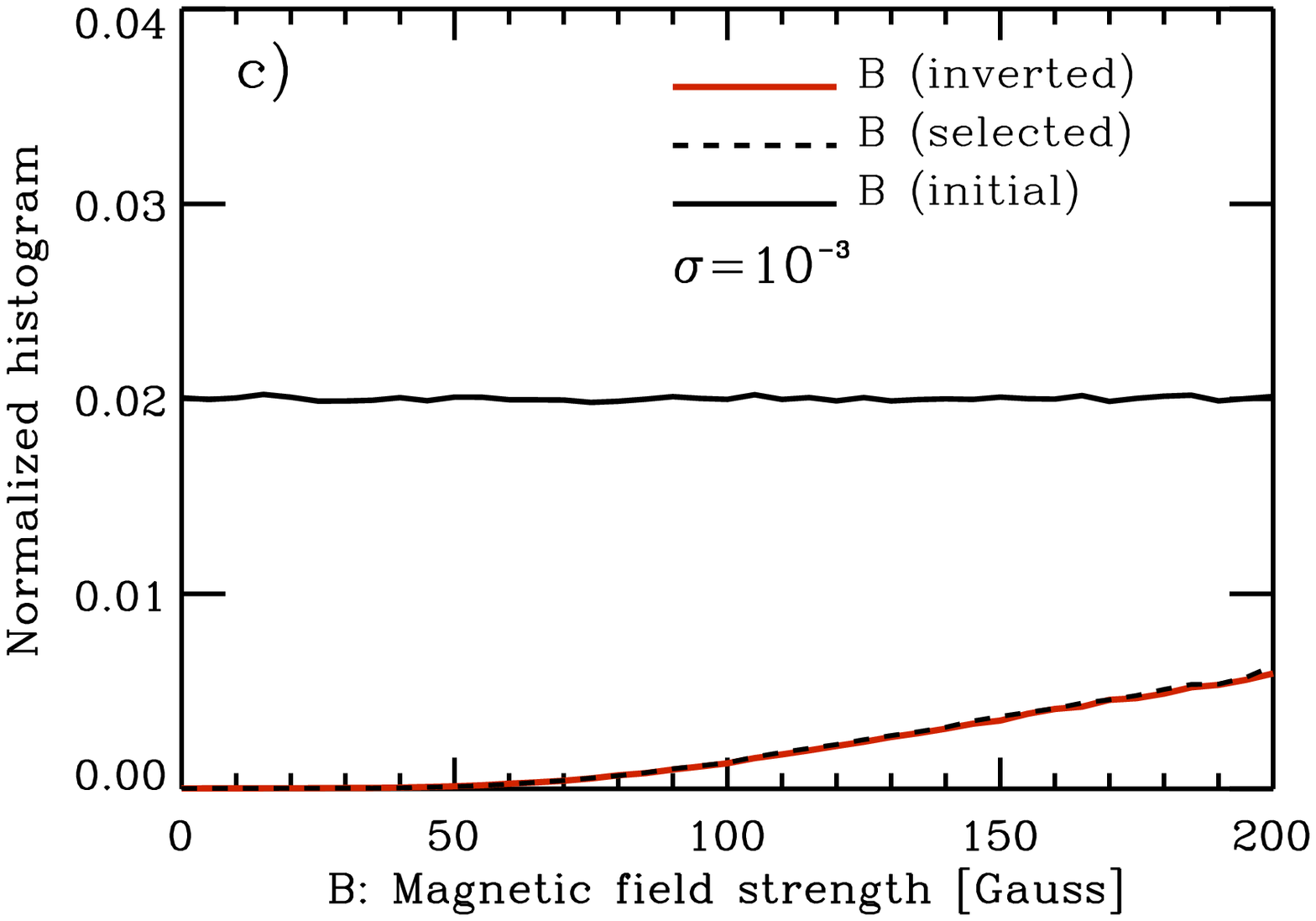} &
\includegraphics[width=8cm]{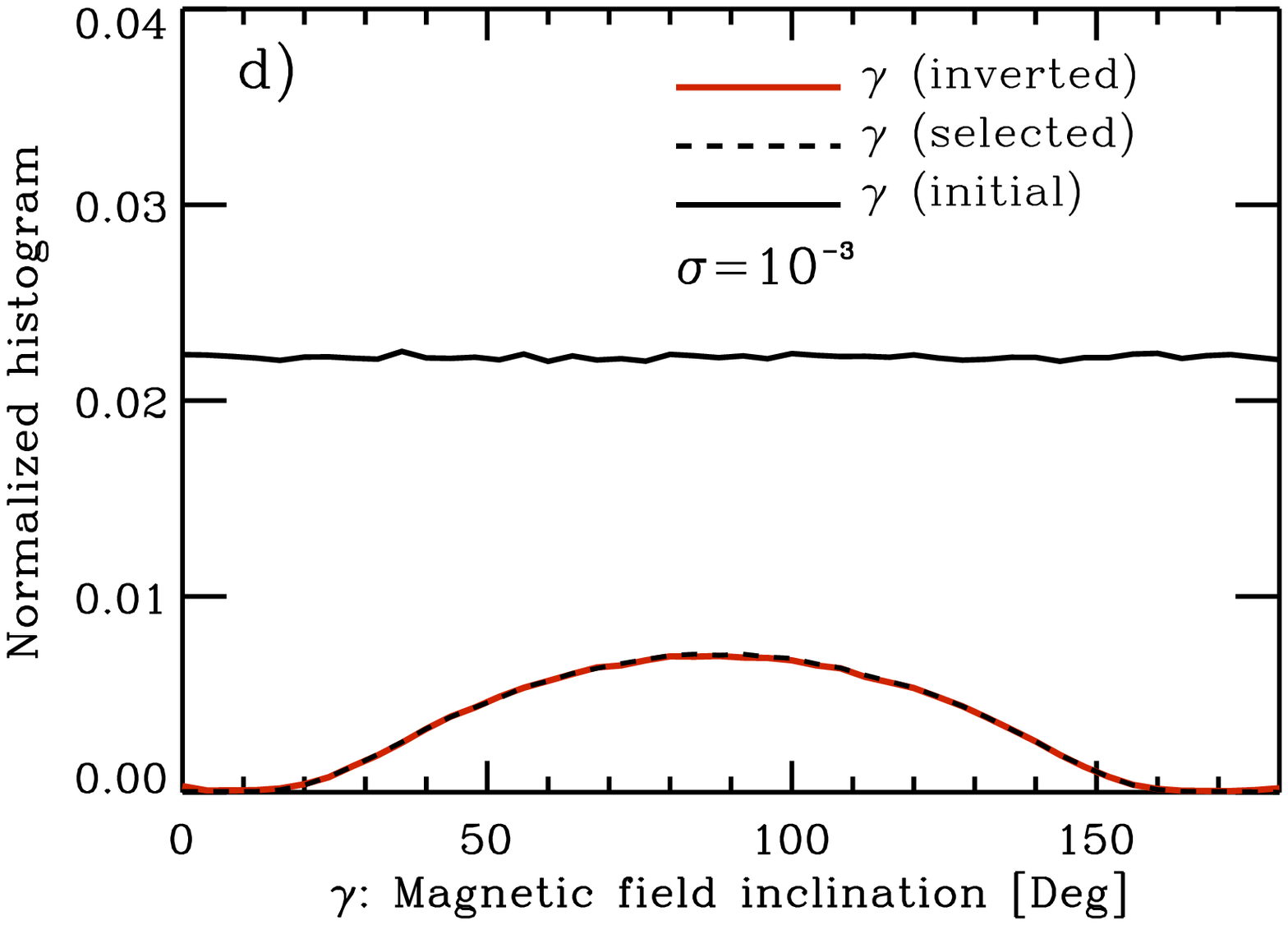} \\
\includegraphics[width=8cm]{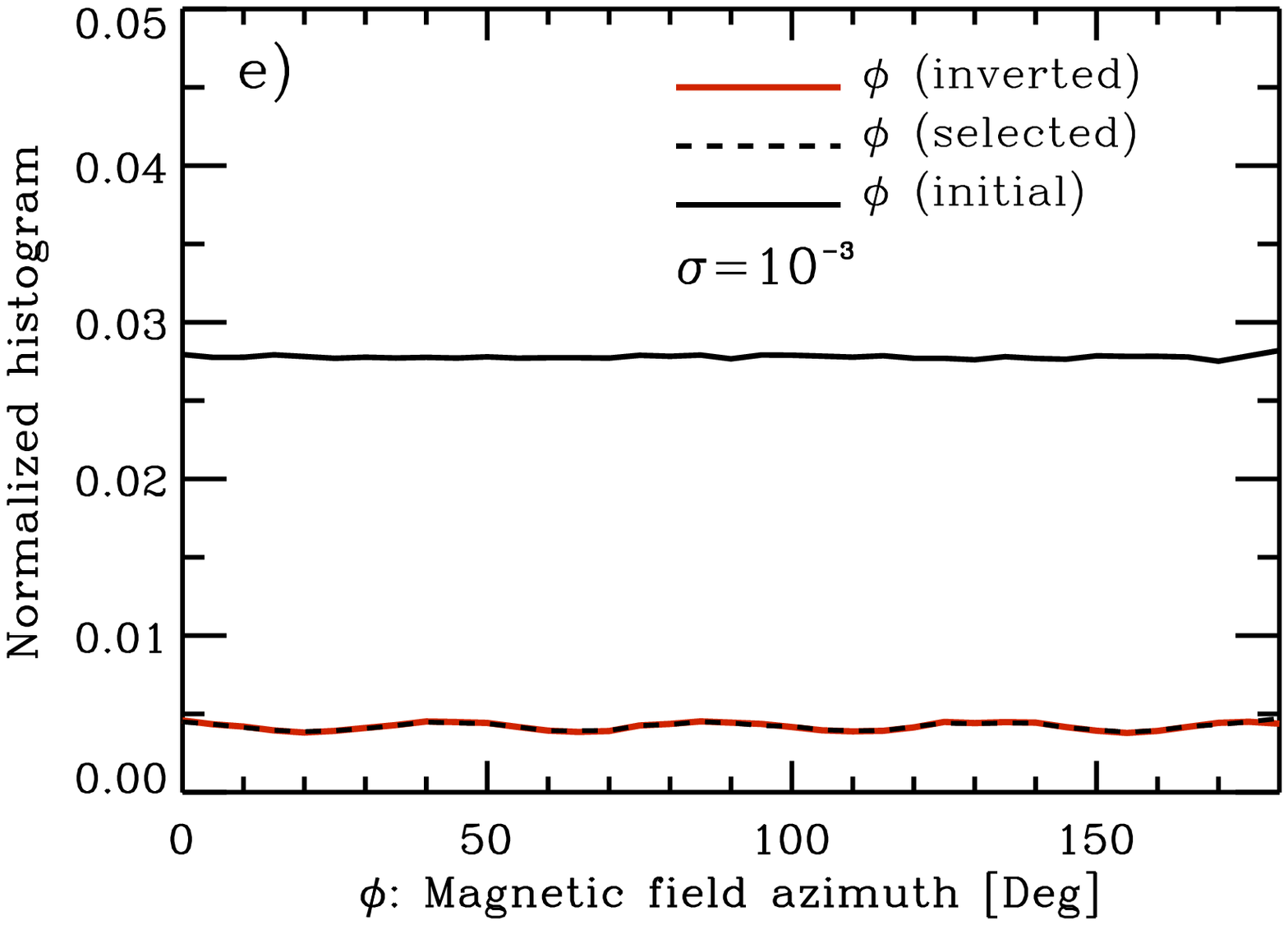} &
\includegraphics[width=8cm]{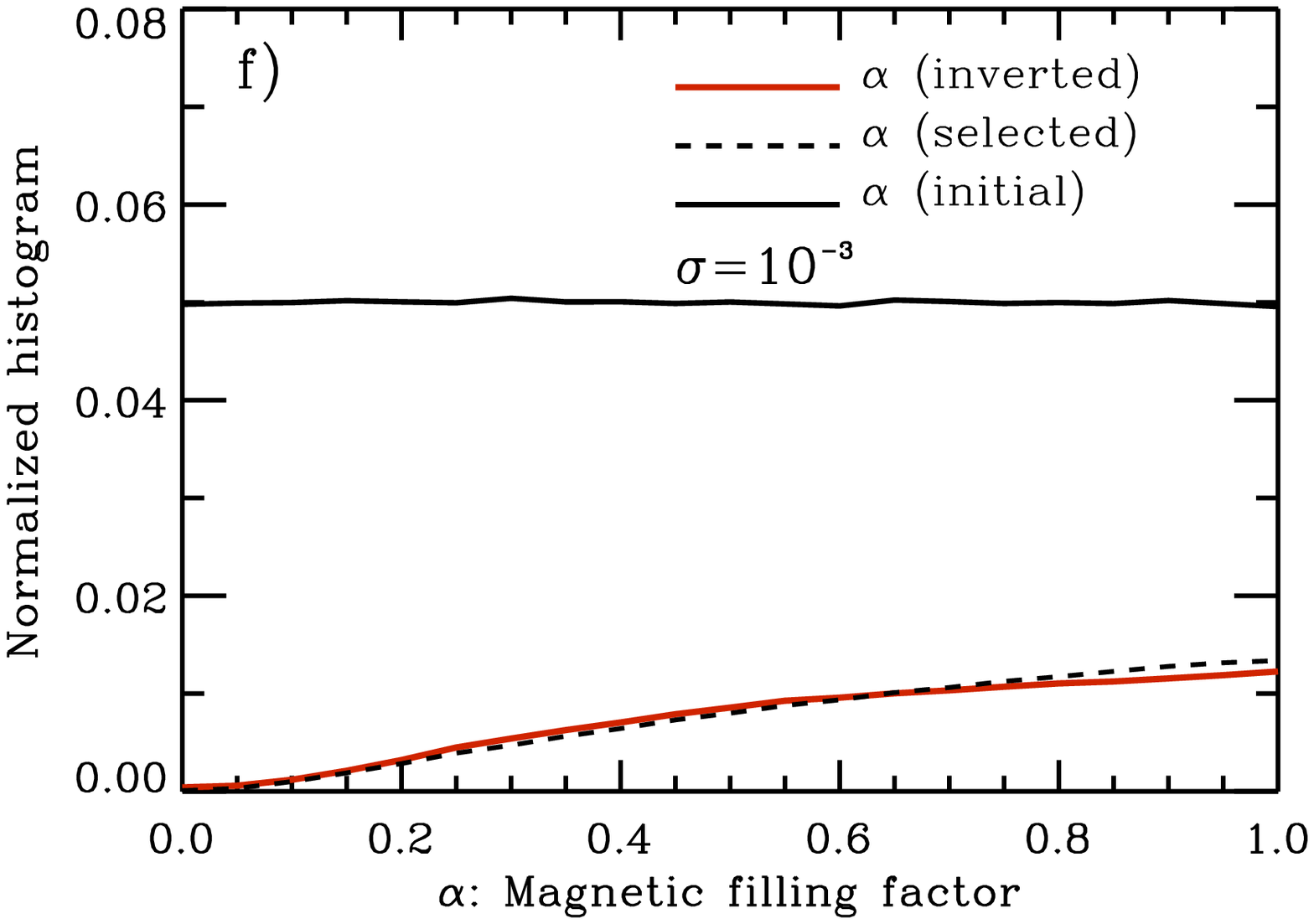} \\
\end{tabular}
\end{center}
\caption{Same as Figure~\ref{figure:highquv} but using a photon noise $\sigma=10^{-3}$ and the $S/R_{\rm qu}$-criterion.}
\label{figure:highqu}
\end{figure*}

\begin{figure*}
\begin{center}
\begin{tabular}{cc}
\includegraphics[width=8cm]{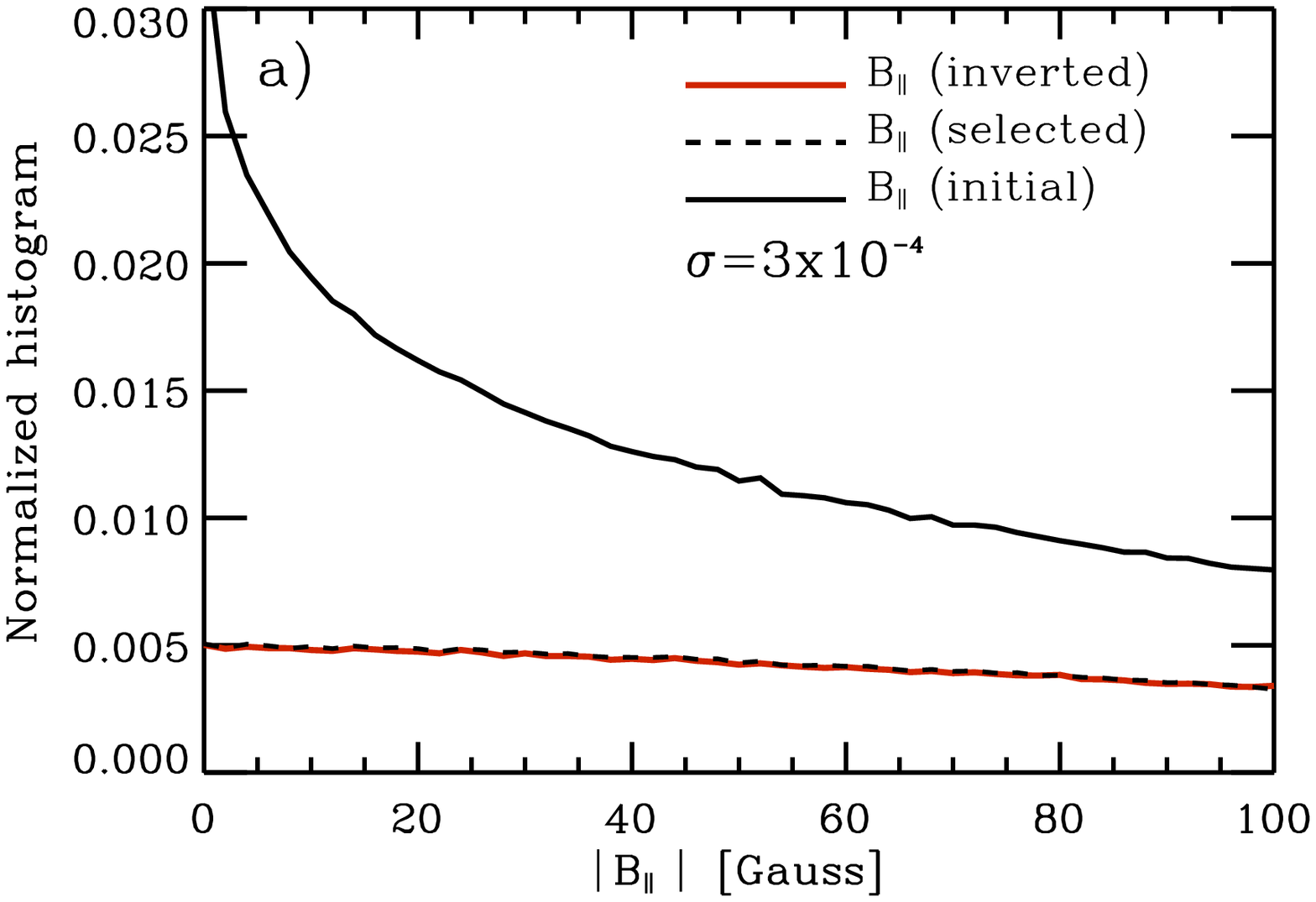} &
\includegraphics[width=8cm]{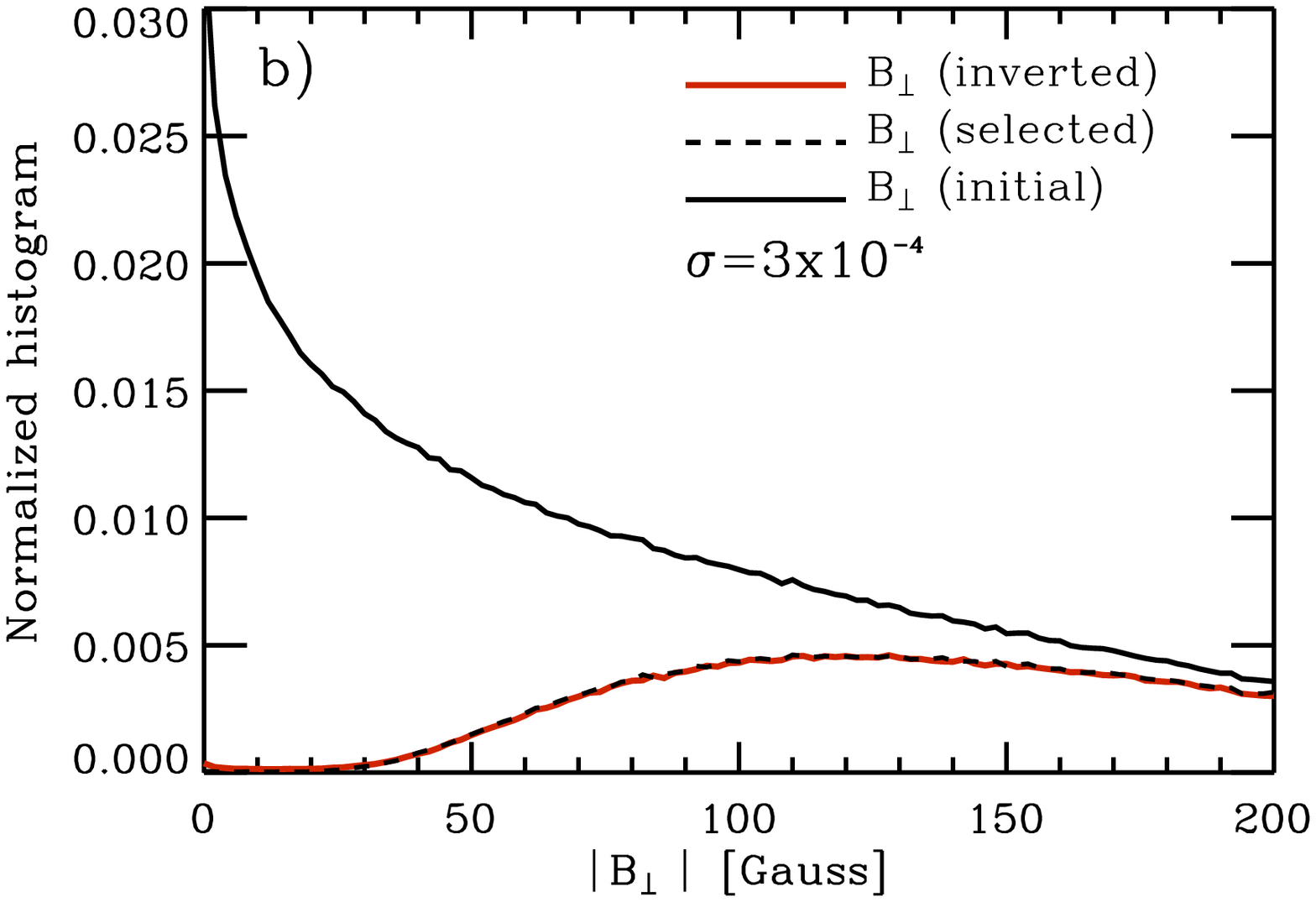} \\
\includegraphics[width=8cm]{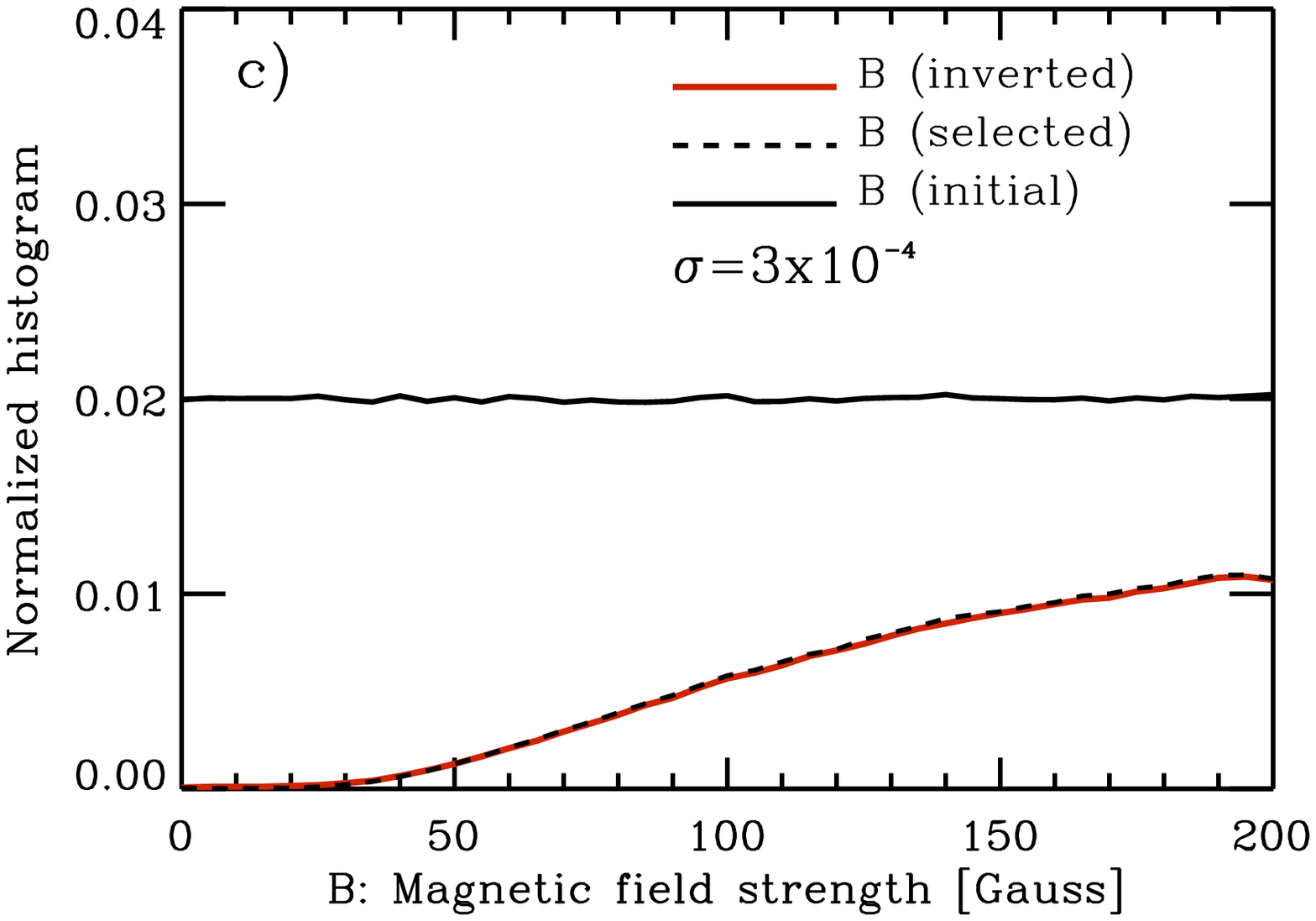} &
\includegraphics[width=8cm]{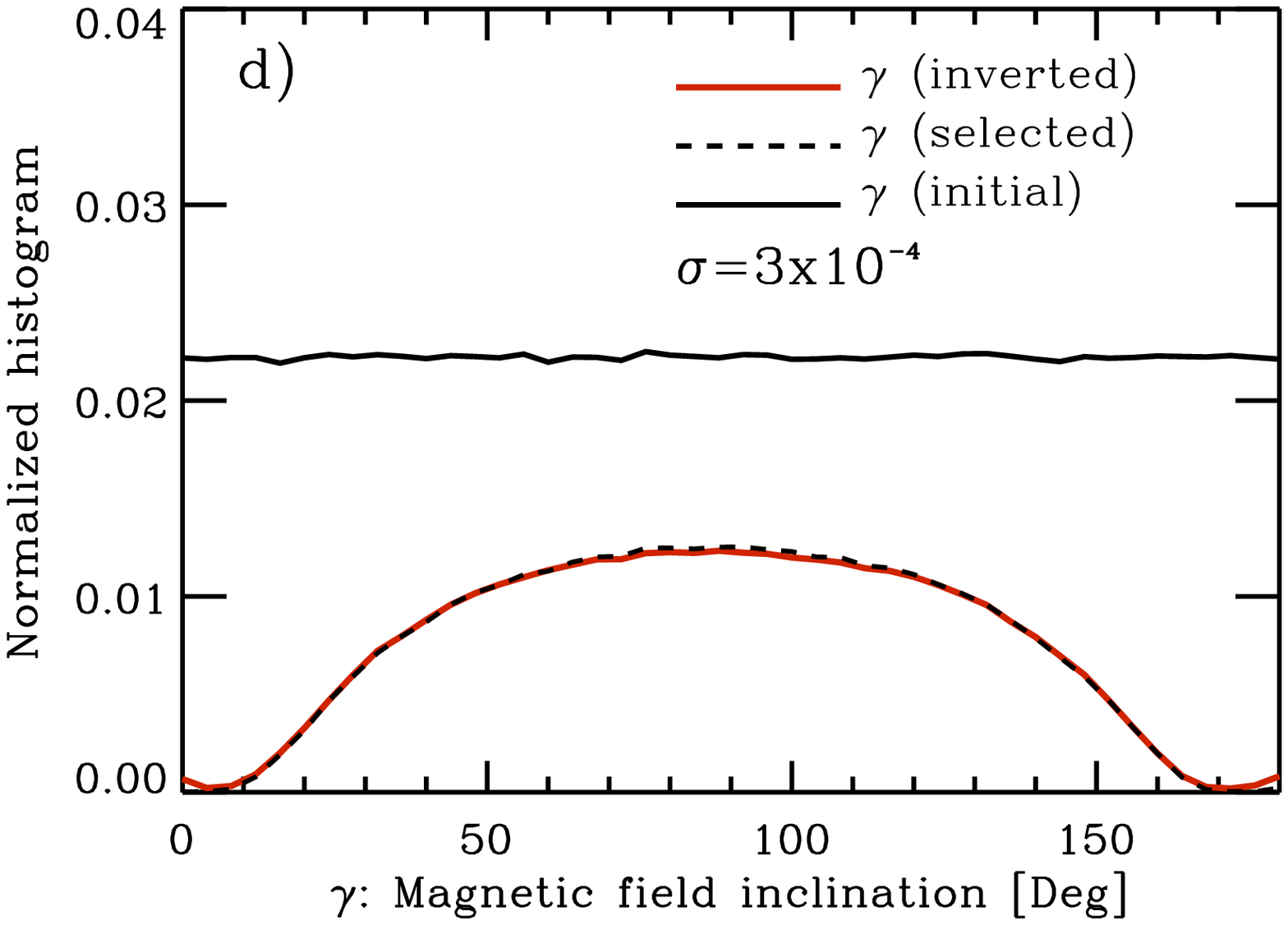} \\
\includegraphics[width=8cm]{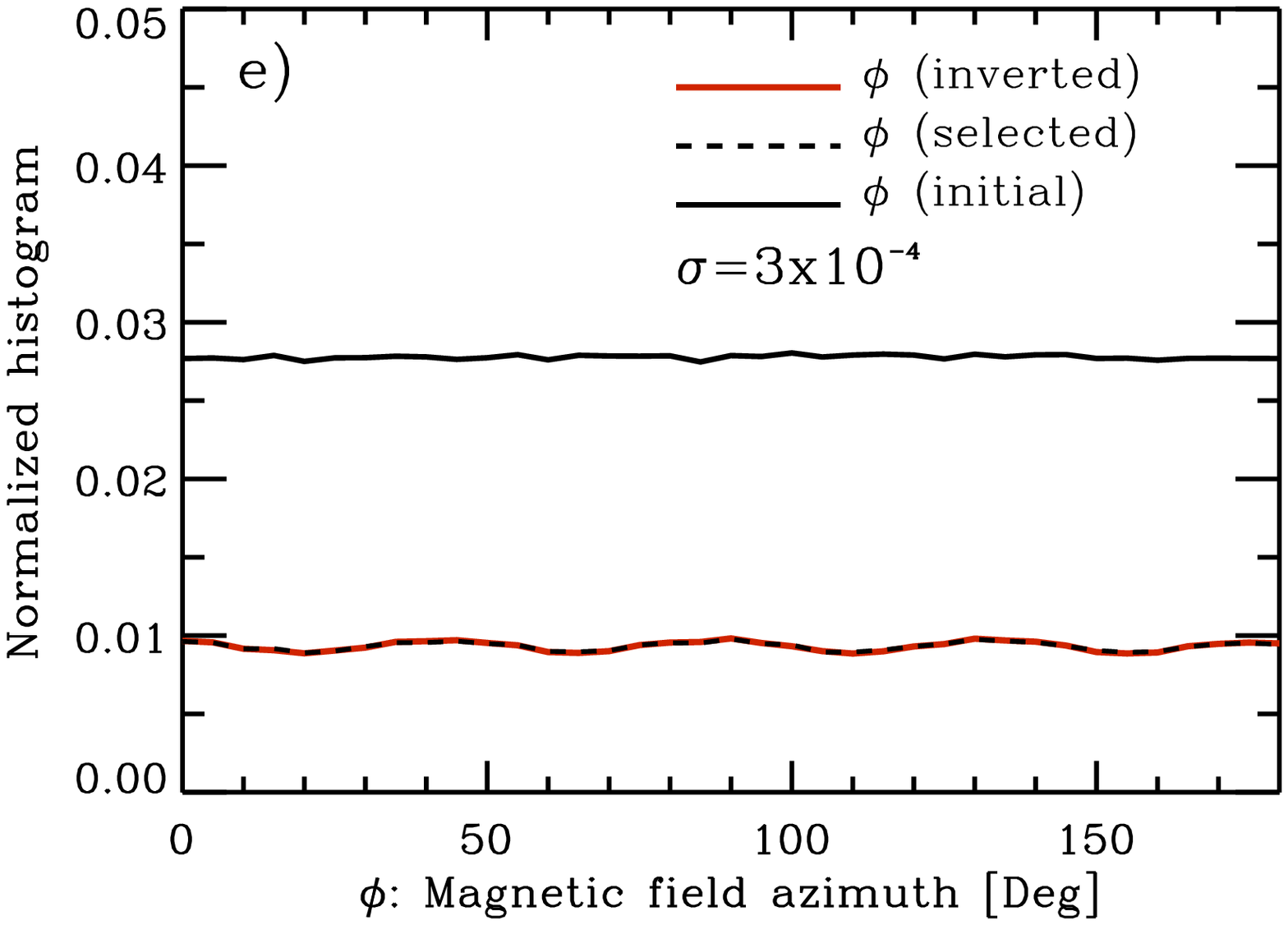} &
\includegraphics[width=8cm]{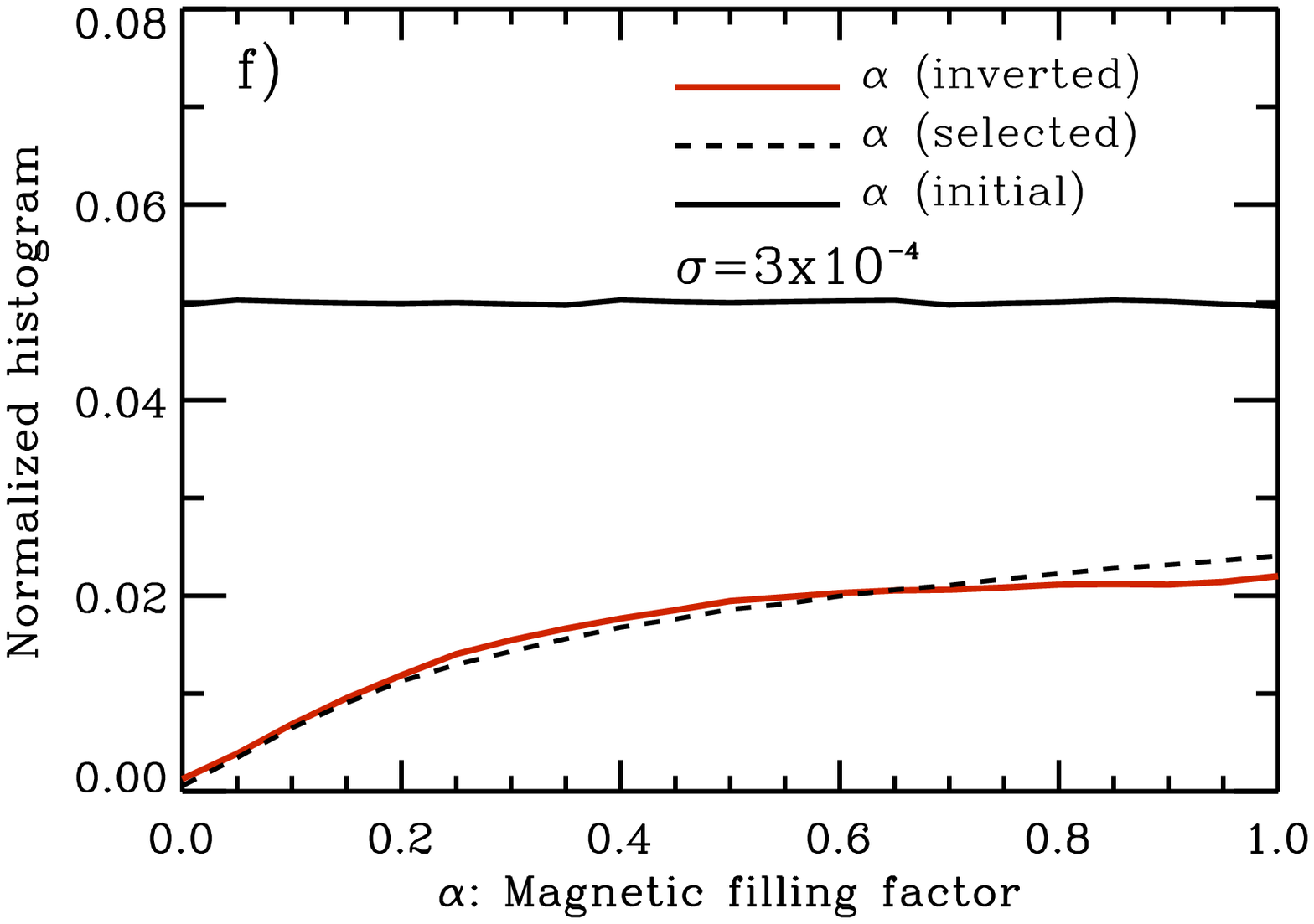} \\
\end{tabular}
\end{center}
\caption{Same as Figure~\ref{figure:highquv} but using a photon noise $\sigma=3\times 10^{-4}$ and the $S/R_{\rm qu}$-criterion.}
\label{figure:lowqu}
\end{figure*}

With the results from the selected profiles using the aforementioned selection criteria, we constructed 
histograms for the magnetic field strength $B$, the inclination of the magnetic 
field vector with respect to the observer's line-of-sight $\gamma$, the projection of the magnetic field
vector along the observer's line-of-sight $B_\parallel = B \cos\gamma$, the projection of the magnetic field vector 
along the perpendicular direction to the observer's line-of-sight: $B_\perp = B \sin \gamma$; azimuthal angle 
of the magnetic field vector in the plane perpendicular to the observer: $\phi$, and finally the magnetic filling factor 
$\alpha$. Results for $\sigma=10^{-3}$ and $S/R_{\rm quv}$ are displayed in Figure~\ref{figure:highquv}, whereas 
Figure~\ref{figure:lowquv} presents the results for $\sigma = 3\times 10^{-4}$ and $S/R_{\rm quv}$. Likewise,
Figures~\ref{figure:highqu} and ~\ref{figure:lowqu} present the results for these same levels of noise, respectively,
but employing the $S/R_{\rm qu}$ selection criterion. The panels in each of these four figures correspond to the following
histograms: {\bf a)} the component of the magnetic field vector along the observer's line-of-sight 
$B_{\parallel}$; {\bf b)} the component of the magnetic field vector perpendicular the observer's line-of-sight 
$B_{\perp}$; {\bf c)} the module of the magnetic field vector $B$; {\bf d)} the inclination of the 
magnetic field vector with respect to the observer's line-of-sight $\gamma$; {\bf e)} the
azimuthal angle of the magnetic field vector in the plane perpendicular to the observer's line-of-sight $\phi$; 
and finally {\bf f)}-panels for the magnetic filling factor $\alpha$. In all panels, solid-black lines represent the
initial distribution employed in the synthesis of the Stokes profiles (see Eqs.~\ref{equation:pdf}-\ref{equation:heaviside}) 
including all $2\times 10^{6}$ Stokes profiles of the distribution. As required, 
the distributions for $B$, $\gamma$, $\phi$, and $\alpha$ are uniform (equal probabilities). We note that this 
is not the case for $B_{\parallel}= B \cos\gamma$ and $B_{\perp} = B \sin\gamma$ because the cosine and sine functions 
introduce non-uniformities in the distribution. In addition, dashed-black lines represent the original distribution
of the physical parameters but considering only the profiles that are selected with corresponding selection criteria,
namely $S/R_{\rm quv}$ in Figs.~\ref{figure:highquv}-\ref{figure:lowquv}, and $S/R_{\rm qu}$ in Figs.~\ref{figure:highqu}-\ref{figure:lowqu}. 
Finally, the solid-red lines present the distributions inferred from the inversion of the Stokes
profiles but employing the same Stokes profiles that were used to construct the dashed-black histograms. In the following we
comment on some general features that will assist us in interpreting our results:\\

\begin{itemize}
\item The area under the solid-black curve is always one, while the area under the dashed-black and solid-red curves are both normalized 
to the quotient of the number of profiles selected with each particular criteria and the total 
number of profiles synthesized ($2\times 10^6$). Therefore, one can interpret the difference between the solid-black and dashed-black 
curves as the loss of information caused by selection criterion, whereas the differences between dashed-black and 
solid-red are caused by the inversion algorithm.\\

\item If the solid-red curve lies below the dash-black one for a given interval of the physical parameter (e.g. $\Delta \gamma$), 
it means that the number of Stokes profiles that were synthesized employing that range of values (e.g. $\Delta \gamma$) are 
underestimated by the inversion. Likewise, whenever the solid-red curve lies above the dash-black one for a given interval, the inversion 
overestimates the number of Stokes profiles that were synthesized employing the range of values of the physical parameter given by the
same interval.\\

\item The area under the dashed-black lines in each panel increases as the noise level decreases. This can be realized by 
comparing Figure \ref{figure:highquv} ($\sigma=10^{-3}$) with Figure \ref{figure:lowquv} ($\sigma=3\times 10^{-4}$), 
and also comparing Figure \ref{figure:highqu} ($\sigma=10^{-3}$) with Figure \ref{figure:lowqu} ($\sigma=3\times 10^{-4}$). 
This happens as a consequence of a larger number of Stokes profiles fulfilling the requirement imposed by a given selection criterion when the 
noise is reduced. Since the $S/R_{\rm qu}$ is more stringent than $S/R_{\rm quv}$ (because linear 
polarization signals are usually much weaker than the circular polarization signals), the area under the dashed-black 
curves is of course much larger in Figs.~\ref{figure:highquv}-\ref{figure:lowquv} than Figs.~\ref{figure:highqu}-\ref{figure:lowqu}.\\
\end{itemize}

\section{Discussion}
\label{section:discussion}

We now discuss qualitatively the effects of photon noise and selection criteria
for each physical parameter individually. We first introduce the approximate
dependences of the Stokes profiles on the three spherical coordinates of the magnetic field vector
($B$, $\gamma$, $\phi$) and the magnetic filling factor $\alpha$, in the weak field approximation
(Landi Degl'Innocenti 1992)\\

\begin{eqnarray}
V & \propto & \alpha B \cos \gamma \notag \\
Q & \propto & \alpha B^2 \sin^2 \gamma \cos 2\phi \\
U & \propto & \alpha B^2 \sin^2 \gamma \sin 2\phi \notag
\label{equation:wfa}
\end{eqnarray}

\subsection{Line-of-sight component of the magnetic field: $B_\parallel$}

As mentioned in the previous Section, the $S/R_{\rm quv}$-criterion selects a portion of the initial Stokes profiles (dashed-black lines in 
Figs.~\ref{figure:highquv}a-\ref{figure:lowquv}a) that more closely resembles the original $B_\parallel$ 
distribution (solid-black lines) than profiles selected by the $S/R_{\rm qu}$-criterion (dashed-black 
lines in Figs.~\ref{figure:highqu}a-\ref{figure:lowqu}a). This is because the 
$S/R_{\rm quv}$-criterion selects a much larger sample of Stokes profiles than $S/R_{\rm qu}$, hence
the distribution obtained with the former criterion must more closely resemble
the original distribution employed to construct the database of Stokes profiles (solid-black lines).\\

We also note that the distribution of the component of the magnetic field vector along the observer's line-of-sight $B_\parallel$ is very 
accurately retrieved by the inversion in all instances, regardless of the selection criterion and either of the noise levels
(see Figs.~\ref{figure:highquv}a-\ref{figure:lowqu}a). This conclusion is reached from the fact that solid-red 
lines coincide almost perfectly with the dashed-black lines in all figures. In addition, the solid-red curves shift
towards lower values of $B_\parallel$ as the noise decreases, which indicates that lower levels of noise allow us to
correctly infer smaller values of the line-of-sight component of the magnetic field.\\

\subsection{Transverse component of the magnetic field: $B_\perp$}

For the same reason given above in the case of $B_\parallel$, the $S/R_{\rm quv}$-criterion selects a 
portion of the initial Stokes profiles (dashed-black lines in Figs.~\ref{figure:highquv}b-\ref{figure:lowquv}b)
that resembles much more closely the original $B_\perp$ distribution (solid-black lines) than profiles selected 
by the $S/R_{\rm qu}$-criterion (dashed-black lines in Figs.~\ref{figure:highqu}b-\ref{figure:lowqu}b). In addition,
the distribution obtained with the $S/R_{\rm qu}$-criterion mostly provides information about the large values of the transverse
component of the magnetic field ($B_\perp \approx 100-150$ G), whereas the $S/R_{\rm quv}$-criterion also selects profiles
arising from weak transverse magnetic fields ($B_\perp \lesssim 50$ G). The reason for this is that much larger
values of $B_\perp$ are needed to produce linear polarization profiles above the $4.5\sigma$-level.\\

The inversion code very reliably retrieves the distribution of $B_\perp$ employing the $S/R_{\rm qu}$-criterion: we note
the almost perfect match between the solid-red curves and dashed-black ones in Figs.~\ref{figure:highqu}b-\ref{figure:lowqu}b.
However, this is clearly not the case for the $S/R_{\rm quv}$-criterion, since here the inversion
underestimates the original distribution in the region where $B_\perp \lesssim 25-40$ G (depending on the noise), but overestimates
it in the region above this threshold. It is also noteworthy that the results from the inversion of the Stokes profiles
selected with the $S/R_{\rm quv}$-criterion show a Maxwellian-like distribution, where the peak appears progressively
at lower values of $B_\perp$ as the noise decreases: $B_\perp^{\rm peak} \approx 60$ G (Fig.~\ref{figure:highquv}b;
$\sigma=10^{-3}$) and $B_\perp^{\rm peak} \approx 35$ G (Fig.~\ref{figure:lowquv}b; $\sigma=3\times 10^{-4}$). This is a 
consequence of having a higher sensitivity for lower levels of noise.\\

\subsection{Total magnetic field strength: $B$}

Since $B=\sqrt{B_\parallel^2+B_\perp^2}$, and since the inversion of Stokes profiles retrieves very reliable 
distributions for $B_\parallel$, any mismatch between the original distributions of $B$ (dashed-black)
and the inferred ones (solid-blue) in Figs.~\ref{figure:highquv}c-\ref{figure:lowqu}c, can be
attributed to the same issues as the inference of $B_\perp$ (see above). In particular, as for
$B_\perp$, the $S/R_{\rm qu}$-criteria allows us to retrieve the distribution of the module of the magnetic field vector $B$ reliably, while
the $S/R_{\rm quv}$-criteria underestimates the real distribution for values $B \lesssim 35-70$ G (depending on the noise) but
overestimates it above this threshold (solid-blue lines in Figs.~\ref{figure:highquv}c and \ref{figure:lowquv}c).\\

\subsection{Inclination of the magnetic field vector: $\gamma$}

As happens with $B_\parallel$ and $B_\perp$, the $S/R_{\rm quv}$-criterion selects a set of Stokes profiles
(dashed-black lines in Figs~\ref{figure:highquv}d-\ref{figure:lowquv}d) that are more representative 
of the initial uniform distribution in $\gamma$ (solid-black lines) than the Stokes profiles chosen 
with the $S/R_{\rm qu}$-criteria (dashed-black lines in Figs~\ref{figure:highqu}d-\ref{figure:lowqu}d).
The $S/R_{\rm quv}$-criterion selects all sorts of inclinations, with a slight preference for more longitudinal
magnetic fields. The $S/R_{\rm qu}$-criterion selects however only highly inclined fields ($\gamma \to 90\deg$). 
This happens because for weak magnetic fields, the only way to have linear polarization signals above the $4.5\sigma$-level
is for these magnetic fields to be highly inclined. This can also be understood in terms of the kind of distributions
obtained in $B_\parallel$ and in $B_\perp$ (panels {\bf a)} and {\bf b)}, respectively) by each selection 
criteria and remembering that $\gamma = \tan^{-1}(B_\perp/B_\parallel)$.\\

By comparing the dashed-black and solid-red lines in Figs.~\ref{figure:highquv}d-\ref{figure:lowqu}d, 
we can conclude that the inversions of the Stokes profiles selected with the $S/R_{\rm quv}$-criterion 
yields a distribution of $\gamma$ that is underestimated for more longitudinal magnetic fields,
but overestimated for more transverse ones. We note that the turning point between the underestimation 
of vertical magnetic fields and the overestimation of horizontal ones occurs at lower values of $\gamma$ as
the noise decreases: $\gamma \simeq 22\deg$ for $\sigma = 10^{-3}$ but only at $\gamma \simeq 17\deg$ for 
$\sigma = 3\times 10^{-4}$. Interestingly, the probability distribution function
around $\gamma = 90\deg$ (where magnetic fields are perpendicular to the line-of-sight) is reliably inferred with this criterion.
Altogether, these findings confirm our previous results (see paper I), for which we concluded that the inversion of the 
Stokes profiles obtained with the $S/R_{\rm quv}$-criterion will correctly infer very inclined magnetic fields, 
whenever these are present, but unfortunately also interpret as very inclined magnetic fields those 
that are mostly aligned with the observer's line-of-sight.  In contrast, the inversion of pixels selected with 
the $S/R_{\rm qu}$-criterion retrieves almost perfectly the distribution of the selected pixels, as can be seen by comparing 
the dashed-black and solid-red curves in Figs.~\ref{figure:highqu}d-\ref{figure:lowqu}d.\\

\subsection{Azimuth of the magnetic field vector: $\phi$}
\label{subsection:azimuth}

Here, both the $S/R_{\rm quv}$ and $S/R_{\rm qu}$-criteria select a set of Stokes profiles that are representative
of the original uniform azimuthal distribution of the magnetic field vector, $\phi$, as the selected
distributions are also very close to uniform (compare the dashed-black and solid-black lines in 
Figures~\ref{figure:highquv}e-\ref{figure:lowqu}e).\\

Remarkably, the inversion of the selected Stokes profiles for both values of the photon noise, $\sigma$,
and both selection criteria is able to retrieve the original distribution of the selected profiles,
as the solid-red lines in Figures~\ref{figure:highquv}e-\ref{figure:lowqu}e match the dashed-black ones.
To a first approximation (Auer et al. 1977; Jeferries \& Mickey 1991), the azimuthal angle of the magnetic field vector 
is given by $\phi = (1/2)\tan^{-1}(U/Q)$. Thus, that the inversion of the Stokes profiles selected 
with the $S/R_{\rm quv}$-criterion also retrieves the correct distribution for $\phi$ comes as a rather surprising result, 
since here most of the selected Stokes $Q$ and $U$ profiles are below the 4.5$\sigma$-level 
(see Section~\ref{section:inversions}). We address this particular point in Section~\ref{section:correlations}.\\

\subsection{Magnetic filling factor: $\alpha$}

In the case of the magnetic filling factor $\alpha$, neither the $S/R_{\rm quv}$
nor the $S/R_{\rm qu}$ criteria (dashed-black lines in Figs.~\ref{figure:highquv}f-\ref{figure:lowqu}f) 
are able to recover correctly the original uniform distribution (solid-black lines). In particular, most of the Stokes
profiles arising from low values of the filling factor, $\alpha \lesssim 0.4$, are neglected
by both selection criteria. This is because the polarization signals scale linearly 
with $\alpha$ (see Equation~\ref{equation:wfa}) and therefore, small values of the magnetic filling factor yield polarization signals
that are below the threshold employed in the selection. The situation is aggravated in the case of $S/R_{\rm qu}$
(Figs.~\ref{figure:highqu}-\ref{figure:lowqu}), where even larger values of $\alpha$ are needed owing
to Stokes $Q$ and $U$ being intrinsically smaller than Stokes $V$.\\

As far as the inversion is concerned, it is clear that the distribution of the magnetic filling factor 
of the selected profiles is very well-retrieved (solid-red lines in Figs.~\ref{figure:highquv}f-\ref{figure:lowqu}f) for both noise 
levels and when employing both selection criteria. Interestingly, as happened with $\gamma$
(although to a much smaller extent), the inversion code slightly overestimates the selected distribution (dashed-black) 
for $\alpha \lesssim 0.5$, but underestimates it above this value.\\

\begin{figure*}
\begin{center}
\begin{tabular}{cc}
\includegraphics[width=8cm]{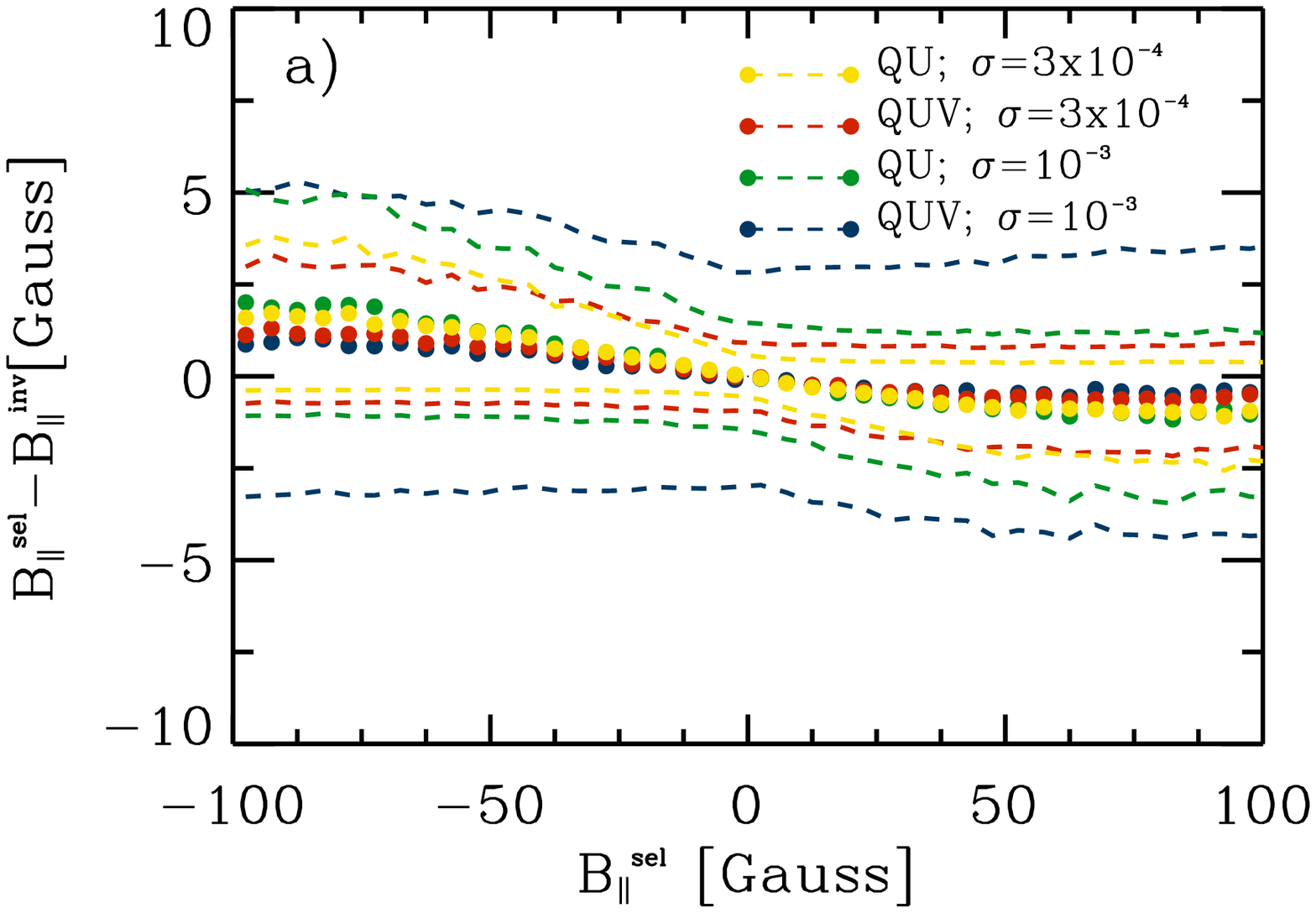} &
\includegraphics[width=8cm]{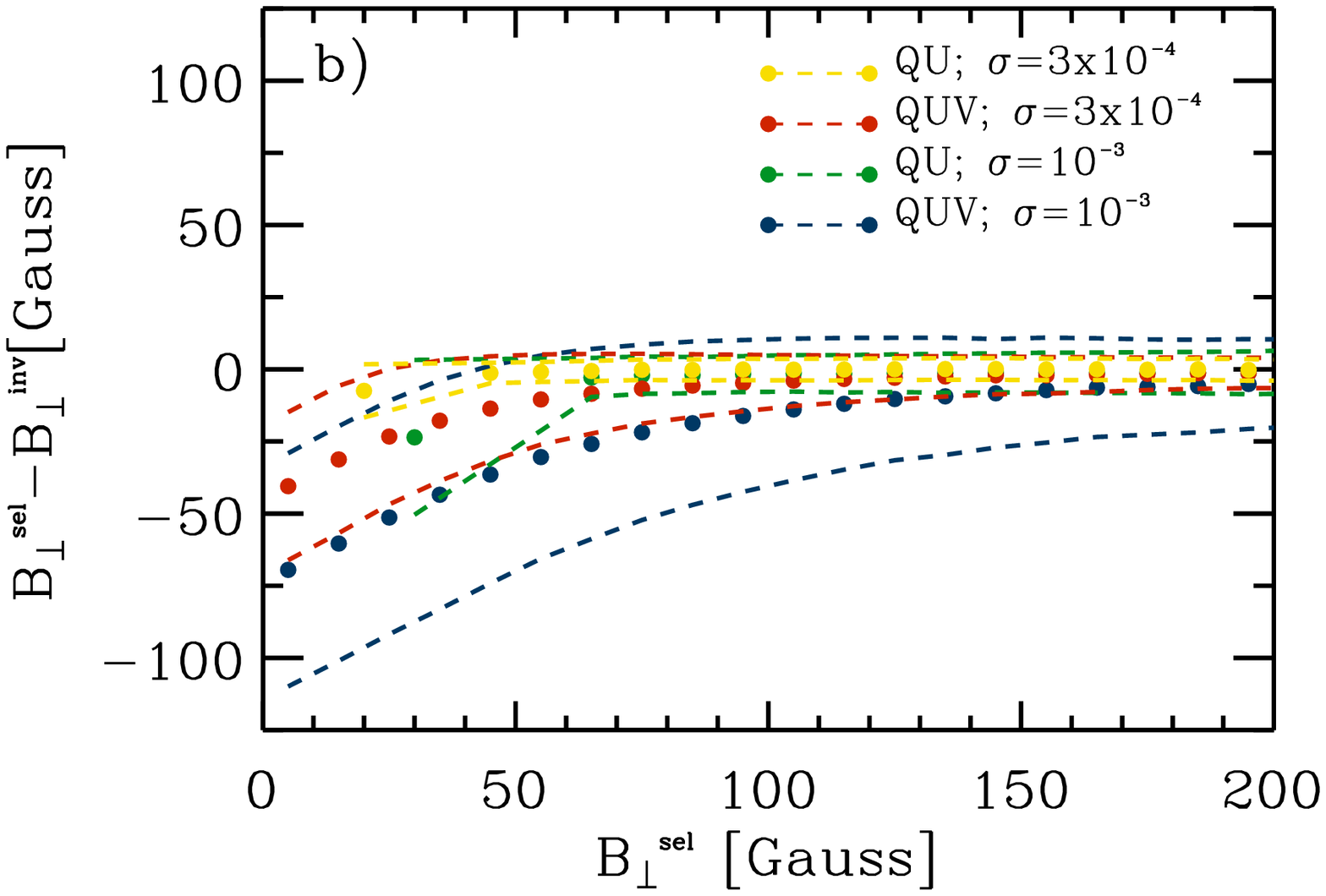} \\
\includegraphics[width=8cm]{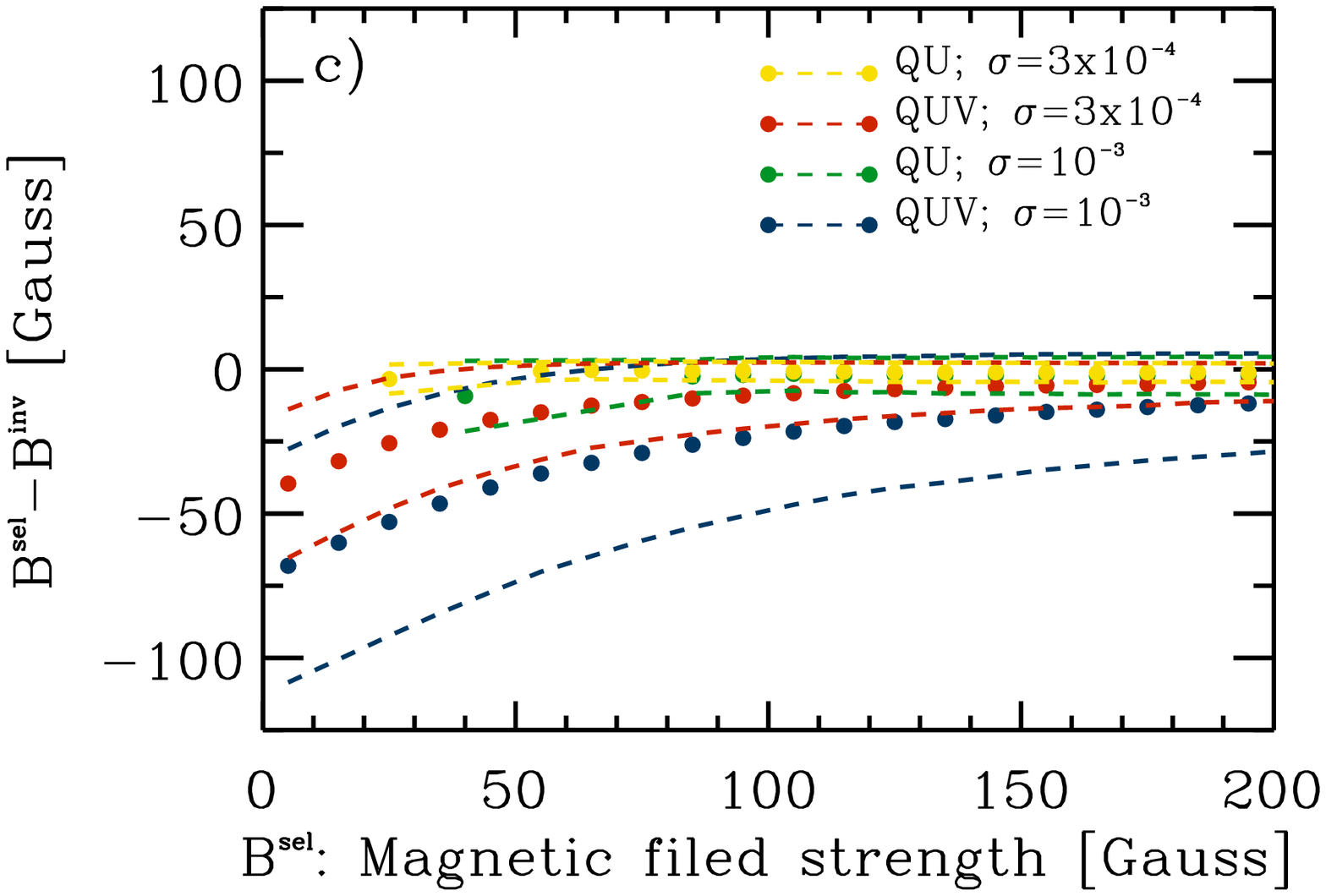} &
\includegraphics[width=8cm]{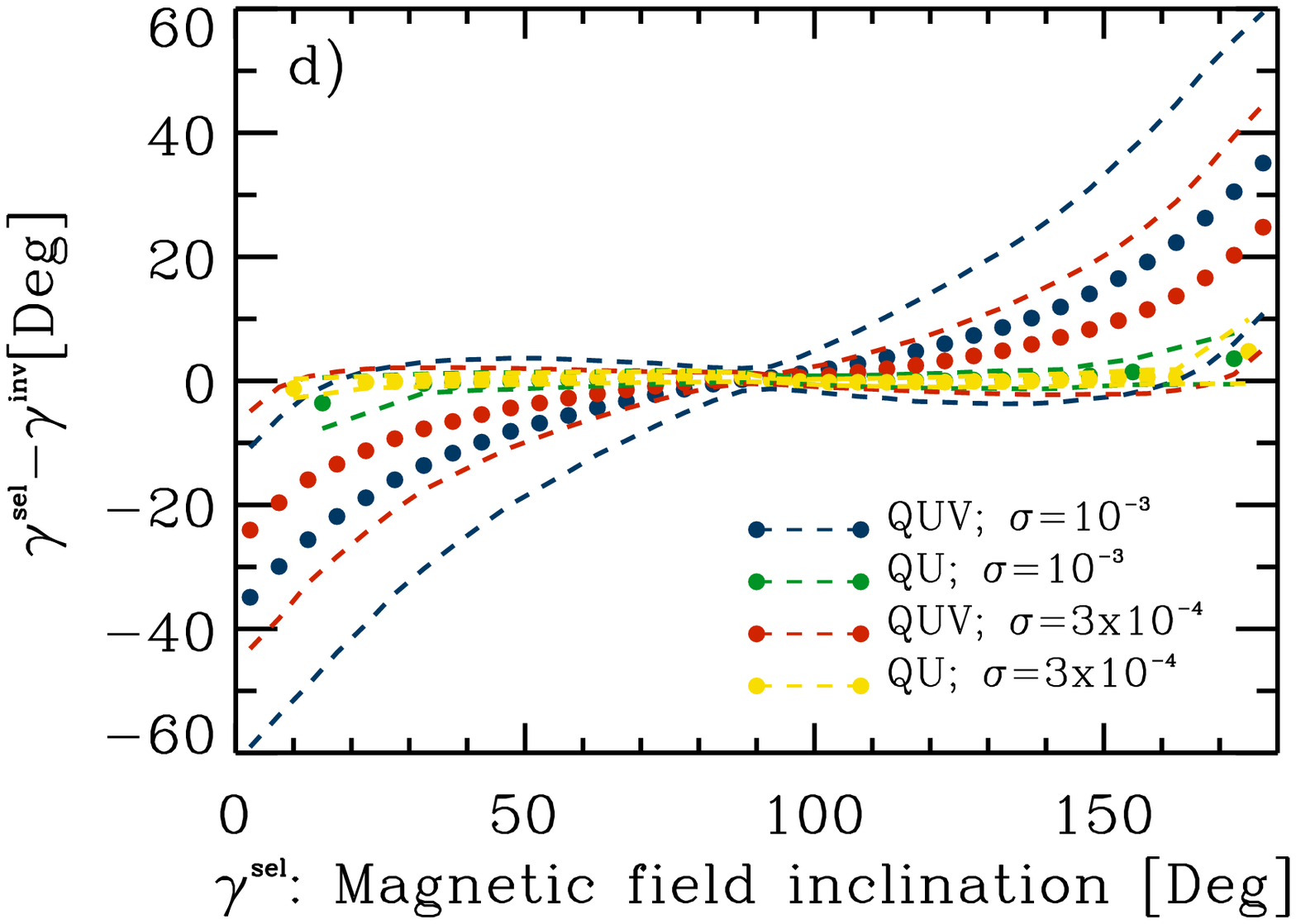} \\
\includegraphics[width=8cm]{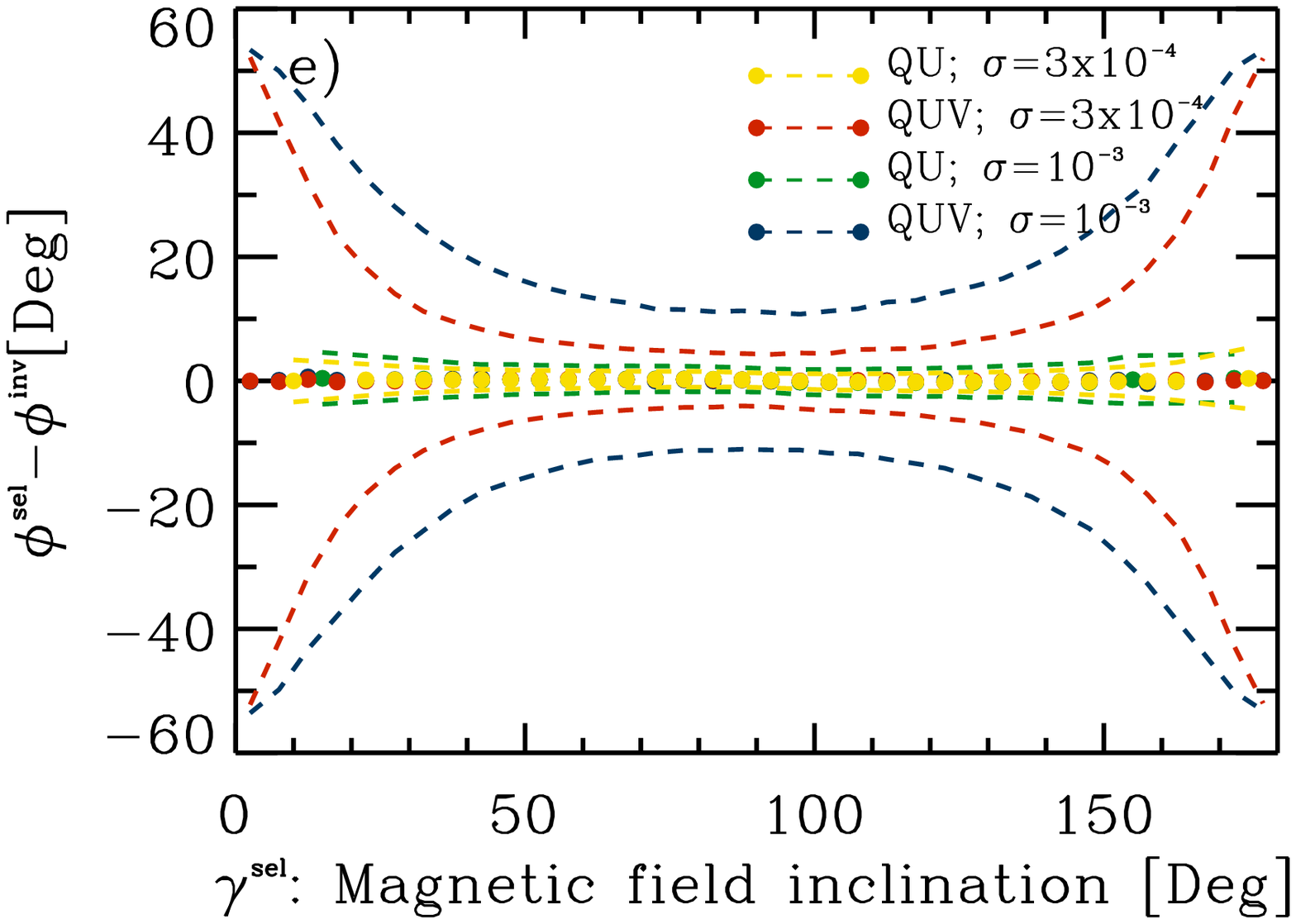} &
\includegraphics[width=8cm]{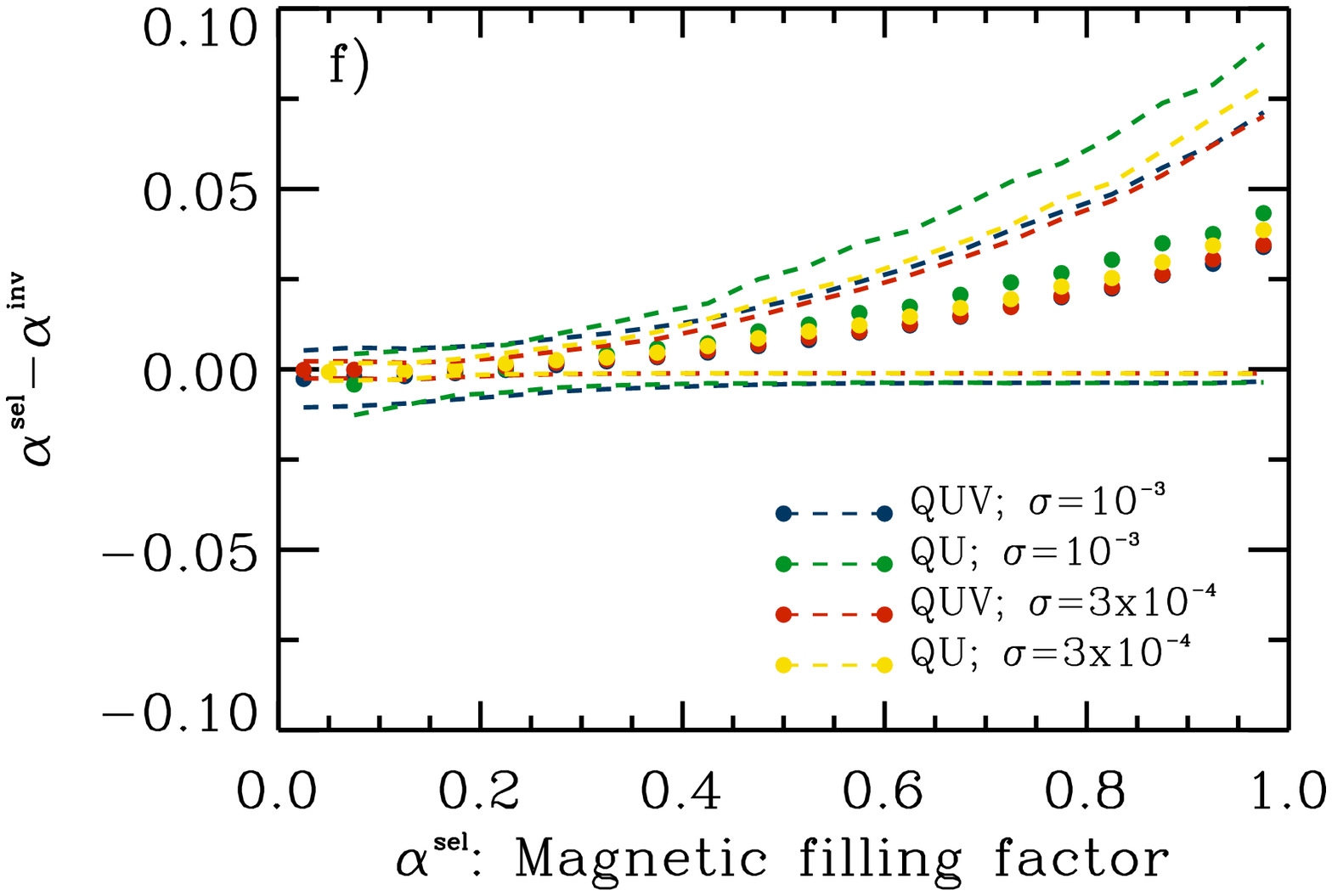} \\
\end{tabular}
\end{center}
\caption{Mean values of the differences between the original physical parameters and the inferred values ($\Delta X_i$; circles)
and standard deviations around the mean (dashed lines; $\sigma_x$): (panel-{\bf a}) $\Delta B_\parallel$ and $\sigma_{B_{\parallel}}$ 
as a function of $B_\parallel$; (panel-{\bf b}) $\Delta B_\perp$ and $\sigma_{B_{\perp}}$ as a function of $B_\perp$; 
(panel-{\bf c}) $\Delta B$ and $\sigma_B$ and a function of $B$; (panel-{\bf d}) $\Delta \gamma$ and $\sigma_\gamma$ as a function 
of $\gamma$; (panel-{\bf e}) $\Delta \phi$ and $\sigma_\phi$ as a function of $\gamma$; and finally, (panel-{\bf f})
$\Delta \alpha$ and $\sigma_\alpha$ as a function of $\alpha$. Blue shows the results obtained from a photon-noise
level of $\sigma=10^{-3}$ and selection criterion $S/R_{\rm quv}$ (from Fig.~\ref{figure:highquv}); red represents $\sigma=3\times 10^{-4}$
and $S/R_{\rm quv}$ (from Fig.~\ref{figure:lowquv}); green is for $\sigma=10^{-3}$ and $S/R_{\rm qu}$ (from Fig.~\ref{figure:highqu});
and yellow is for $\sigma=3\times 10^{-4}$ and $S/R_{\rm qu}$ (from Fig.~\ref{figure:lowqu}).}
\label{figure:errors}
\end{figure*}

We have so far only discussed the ability of the inversion code to retrieve the correct distribution 
for the magnetic parameters (three components of the magnetic field vector and filling factor). Although this is
clearly an important question, it does not provide much information about the reliability of the inversion
in individual cases. To address this point, we display in circles in Figure~\ref{figure:errors}, the mean value of the
differences between the original physical parameters of the selected Stokes profiles and the inferred
ones through the inversion $X^{\rm sel}-X^{\rm inv}$. These are denoted as $\Delta X_i$, with $X_i$ being any of the components of $\ve{X}$
in Equation~\ref{equation:x}, which we refer to as \emph{bias in $X_i$}. In addition to this, we also plot the \emph{standard deviation
around the mean} in the inference of the physical parameters as the dashed-color lines, which we refer to as $\sigma_x$ (not to be 
confused with the photon noise $\sigma$). For a physical parameter to be well-constrained, it is important that the mean is centered around zero 
(otherwise a systematic bias occurs) and that the standard deviation is small.\\

We note that the size of the intervals in which the bias and standard deviation are calculated is not constant. This happens as a 
consequence of the lack of profiles in certain ranges. For instance, the $S/R_{\rm qu}$ criterion barely selects profiles arising 
from a magnetic field vector where $B_\perp < 50$ G (see red and yellow circles in Figs.~\ref{figure:highqu}b-\ref{figure:lowqu}b). 
This makes it necessary to increase the interval of $B_\perp$ (around this range) so that a sufficient number of profiles can 
be employed to obtain a statistically meaningful $\Delta B_\perp$ and $\sigma_{B_\perp}$ (see green and yellow circles in 
Figure~\ref{figure:errors}b). This is not needed as often in the $S/R_{\rm quv}$-criterion because it selects many more
profiles than the $S/R_{\rm qu}$-criterion (blue and red circles in Fig.~\ref{figure:errors}b).\\

By inspection of Figure~\ref{figure:errors}, we note that, for the same level of noise and with the exception of the
magnetic filling factor $\alpha$, the standard deviation in the retrieval of the physical parameters is always much 
smaller when employing the $S/R_{\rm qu}$-criterion than the $S/R_{\rm quv}$-criterion. By comparing the standard deviations
with different levels of photon noise, we realize that for most physical parameters, the improvement achieved by a decrease in the 
photon noise (from $\sigma=10^{-3}$ to $3\times 10^{-4}$) is smaller than the improvement achieved by using the $S/R_{\rm qu}$ instead of 
the $S/R_{\rm quv}$-criterion. This, generally makes the dashed-green curves ($\sigma=10^{-3}$ and $S/R_{\rm qu}$-criterion) lie below the red-dashed ones 
($\sigma=3\times 10^{-4}$ and $S/R_{\rm quv}$-criterion). Interestingly, this does not apply to all physical parameters: in the case of 
$B_\parallel$ and $\alpha$, the photon noise plays a more important role than the selection criteria itself, as the red curves lie below the green ones.
The relative importances of the selection criteria and the photon noise logically depends on the values of the noise considered,
thus we cannot conclude that in general the former is more important than the latter (or the other way around) for some particular 
physical parameter.\\

We now discuss the errors in the case of the $S/R_{\rm quv}$-criterion. Here, the retrieval of 
$B_\perp$ (blue and red colors in Fig.~\ref{figure:errors}b) has a large systematic bias towards 
$B_\perp^{\rm inv} > B_\perp^{\rm sel}$, such as $\Delta B_\perp < -75$ G, for $B_\perp^{\rm sel} \lesssim 50$ G.
Furthermore, in this same region, the standard deviation is as large as $\sigma_{B_\perp} \approx 20-50$ G. These numbers decrease 
as $B_\perp^{\rm sel}$ increases, such that $\Delta B_\perp \approx -15$ G and $\sigma_{B_\perp} \approx 10$ G for $B_\perp^{\rm sel} > 150$ G. 
As happened in the case of the distributions in Figs.~\ref{figure:highquv}c-\ref{figure:lowqu}c, the bias and standard deviations in the module of the magnetic field 
vector $B$ (Figure~\ref{figure:errors}c) mimic those of $B_\perp$ (Figure~\ref{figure:errors}b). In the case of $B_\parallel$ (blue and red colors in 
Fig.~\ref{figure:errors}a), both the bias and standard deviation are always quite small $< 5$ G, owing to the Stokes $V$ 
profiles typically being much larger than $Q$ and $U$ for the same values of $B_\parallel$ and $B_\perp$. For the inclination 
of the magnetic field vector with respect to the observer's line-of-sight, $\gamma$ (blue and red colors in Fig.~\ref{figure:errors}d), 
the bias is systematic towards more inclined magnetic fields: $\Delta \gamma < 0$ if $\gamma^{\rm sel} < 90\deg$, but $\Delta \gamma > 0$ if $\gamma^{\rm sel} > 90\deg$.
The absolute value of the bias is indeed larger for magnetic fields aligned with the observer's line-of-sight, where $| \Delta \gamma | \approx 30-40\deg$ 
if $\gamma^{\rm sel} \approx 0, 180\deg$. Interestingly, the bias decreases as the magnetic field vector becomes more perpendicular 
to the line-of-sight, i.e., $\Delta \gamma= 0$ for $\gamma^{\rm sel} \approx 90\deg$. The behavior of $\sigma_\gamma$
is very similar (at all values) to that of the absolute values of bias $| \Delta \gamma |$. That the bias and standard deviation 
decrease as $\gamma \to 90\deg$ indicates that when the magnetic field vector is not completely aligned with the observer's line-of-sight, the signal
that appears in $Q$ and $U$ helps the inversion code to extract some information about the inclination of the magnetic field vector. This inference 
is however negatively affected by the photon noise systematically overestimating $\gamma$. This agrees with our results 
in paper I, where we found that it was impossible to distinguish between a vertical and horizontal magnetic-field vector when Stokes $Q$ and $U$ 
are below the 4.5$\sigma$-level. Figure~\ref{figure:errors}e displays $\Delta \phi$ and $\sigma_\phi$ as a function of $\gamma^{\rm sel}$ instead 
of $\phi^{\rm sel}$. In this case, the bias in the determination of the azimuthal angle of the magnetic field vector is almost negligible for 
all possible values of $\gamma^{\rm sel}$, where $\Delta \phi \approx 0\deg$. However, the standard deviation
is large ($\sigma_\phi \approx 40-50\deg$) when the magnetic field vector is aligned with the observer's line-of-sight ($\gamma \to 0,180\deg$), but decreases
as the magnetic field vector becomes more and more inclined with respect to the observer's line-of-sight ($\gamma \to 90\deg$). 
This happens as a consequence of the linear polarization profiles vanishing when the magnetic field becomes aligned with the observer such that
$Q,U \propto \sin^2\gamma$ (see Eq.~\ref{equation:wfa}).\\

Compared to the $S/R_{\rm quv}$-criterion, there is no bias in the retrieval of all parameters (except for the magnetic filling factor) 
when employing the $S/R_{\rm qu}$-criterion $\Delta X_i \to 0$. In addition, the standard deviation in the retrieval 
of the physical parameters, $\sigma_x$, greatly decreases when employing the $S/R_{\rm qu}$-criterion,
as indicated by the green and yellow colors in Figure~\ref{figure:errors}. With this criterion $\Delta B_\perp
\approx 0$ G and $\sigma_{B_\perp} \approx 10$ G, even for $B_\perp < 50$ G. The retrieval of $B_\parallel$ also improves,
(such that $\sigma_{B_{\parallel}} < 3$ G) from that achieved with the $S/R_{\rm quv}$-criterion. Finally, this translates into smaller standard deviations
for $B$, $\gamma$, and $\phi$: $\sigma_B < 5 $ G, $\sigma_\gamma < 2\deg$, and $\sigma_\phi < 2\deg$.\\

\section{Can inversion codes retrieve the distribution of the orientation of 
the magnetic field vector from correlations within the noise ?}
\label{section:correlations}

\subsection{Azimuth of the magnetic field vector: $\phi$}
\label{subsection:phinoise}

\begin{figure}
\begin{center}
\includegraphics[width=9cm]{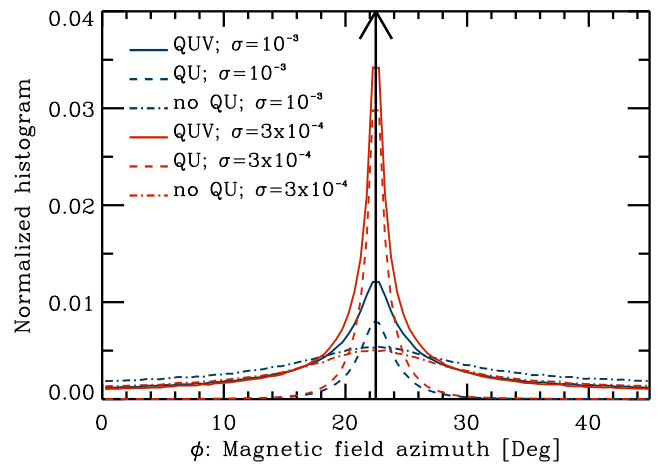} \\
\end{center}
\caption{Histograms of the inferred azimuthal angle of the magnetic field vector, $\phi$, when employing Equation~\ref{equation:pdfdeltaazim}
in the synthesis of Stokes profiles. The original distribution $\delta(\phi-\pi/8)$ is indicated by the vertical black arrow located at $\phi=22.5\deg$.
The inferred histograms using $S/R_{\rm quv}$, $S/R_{\rm qu}$, and $S/R_{\rm no-qu}$ criteria are indicated by the solid, dashed, and dotted-dashed
lines, respectively. Blue curves are obtained with a photon noise $\sigma=10^{-3}$, whereas red ones correspond to $\sigma=3\times 10^{-4}$.}
\label{figure:azimuth}
\end{figure}

In Section~\ref{subsection:azimuth}, we have seen that the retrieved distribution of the azimuthal angle of the magnetic field vector
on the plane that is perpendicular to the observer's line-of-sight ($\phi$) matches excellently well the original 
uniform distribution for the selected pixels. In the case of the $S/R_{\rm quv}$-criterion, this seems counter-intuitive since
many of the selected Stokes profiles only have sufficient signal in Stokes $V$, while $Q$ and $U$ are usually dominated by photon noise.
One possible explanation for this is that the inversion does not properly infer $\phi$ but instead yields a random distribution of values 
owing to the lack of information in the linear polarization profiles. If this distribution
happens to be uniformly distributed, then the inferred distribution of azimuths matches the original one, even if the 
inversion fails for each Stokes vector individually. The second explanation is that, given that the number of profiles with large enough
signals in Stokes $Q$ and $U$ is non-negligible, the inversion indeed properly retrieves $\phi$. In particular, in Section~\ref{section:inversions} 
we mentioned that 21.2 \% ($\sigma=10^{-3}$) and 39.2 \% ($\sigma=3\times 10^{-4}$) of the profiles selected with the $S/R_{\rm quv}$-criteron, 
had linear polarization signals above the $4.5\sigma$-level. A final possible explanation is that the inversion is able to retrieve 
the correct values of $\phi$ even when Stokes $Q$ and $U$ are dominated by noise. This could happen if
the hidden signal introduces a sufficiently strong correlation in the noise, such that the minimization process can properly
retrieve the azimuth of the magnetic field vector. To investigate which of these three possibilities are responsible
for our results in Section~\ref{section:discussion}, we repeated the synthesis (Sect.~\ref{section:uniformpdf})
and inversion experiments of the spectral line Fe I 6302.5 {\AA}, but employing the probability distribution function\\

\begin{eqnarray}
\mathcal{P}_1(\ve{B},\alpha) \df\ve{B} \df\alpha = \frac{1}{\pi B_0} \delta\left(\phi-\frac{\pi}{8}\right) \textrm{H}(B-B_0) \df B \df\gamma \df\phi 
\df\alpha \;,
\label{equation:pdfdeltaazim}
\end{eqnarray}

\noindent which is very similar to Equation~\ref{equation:pdfuniform} with the exception that the distribution of
the azimuthal angle is now a $\delta$-Dirac centered at 22.5$\deg = \pi/8$ rad. This value is chosen such that
$Q \propto \sin^2\gamma\cos 2\phi$, and $U \propto \sin^2\gamma\sin 2\phi$, have the same amplitudes (see Eq.~\ref{equation:wfa}). 
The distribution given by Equation~\ref{equation:pdfdeltaazim} is employed to produce synthetic Stokes profiles, which are then 
inverted after adding photon noise with levels of $\sigma = 10^{-3}$ and $3\times 10^{-4}$. The resulting distributions, for 
both levels of noise and both selection criteria ($S/R_{\rm quv}$ and $S/R_{\rm qu}$), are shown in Figure~\ref{figure:azimuth}.\\

Figure~\ref{figure:azimuth} shows that, on the one hand, the $S/R_{\rm quv}$-criterion (solid lines) has an almost uniform background 
distribution. This is caused by the inversion of Stokes profiles whose linear polarization signals are so weak that 
they carry no information about the azimuth, thus the inversion retrieves random results that happen to be uniformly 
distributed. We know this because this background uniform field disappears if we apply the $S/R_{\rm qu}$-criterion (dashed lines), 
which selects only Stokes $Q$ and $U$ profiles that are 4.5 times above the noise level. This is therefore one of the reasons for the
excellent results in Figures~\ref{figure:highquv}e-\ref{figure:lowquv}e.\\

In Figure~\ref{figure:azimuth}, the $S/R_{\rm quv}$-criterion also features (superimposed on the uniform background distribution) 
a clear reminiscent signature of the original $\delta(\phi-\pi/8)$ distribution (indicated by the vertical black arrow). This indicates that
the inversion code is able to retrieve, to some extent, information about the azimuthal angle of the magnetic field vector. However,
we still do not know whether $\phi$ is correctly retrieved in only 21.2/39.2 \% ($\sigma=10^{-3},3\times 10^{-4}$, respectively) 
of the Stokes profiles that have sufficient signal in the linear polarization or, whether the inversion can properly infer $\phi$
even when the linear polarization signals are below the 4.5$\sigma$-level. To answer this question, we devised a third selection
criterion, $S/R_{\rm no-qu}$, for which we selected all profiles where the linear polarization signals were below the $4.5\sigma$-level.
The distribution of $\phi$ obtained with this criterion (dotted-dashed lines in Figure~\ref{figure:azimuth}) shows an almost uniform
distribution, with a far smaller peak at $\phi=\pi/8$ than before. This indicates that when the signals are dominated by noise, the correlation hidden 
below the noise does not provide sufficient information about the azimuthal angle of the magnetic field vector.\\

\subsection{Inclination of the magnetic field vector: $\gamma$}
\label{subsection:gammanoise}

\begin{figure}
\begin{center}
\includegraphics[width=9cm]{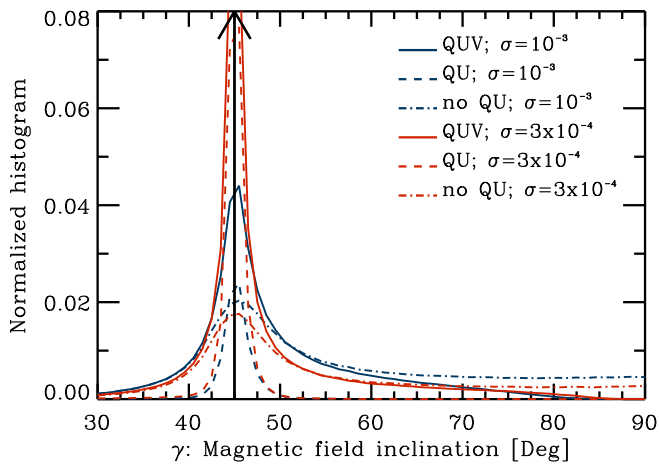} \\
\end{center}
\caption{Histograms of the inferred inclination angle of the magnetic field vector, $\gamma$, when employing Equation~\ref{equation:pdfdeltagamma}
in the synthesis of Stokes profiles. The original distribution $\delta(\gamma-\pi/4)$ is indicated by the vertical black arrow located at $\phi=55\deg$.
The inferred histograms using $S/R_{\rm quv}$, $S/R_{\rm qu}$, and $S/R_{\rm no-qu}$ criteria are indicated by the solid, dashed, and dotted-dashed 
lines, respectively. Blue curves are obtained with a photon noise $\sigma=10^{-3}$, whereas red ones correspond to $\sigma=3\times 10^{-4}$.}
\label{figure:gamma}
\end{figure}

In paper I, we demonstrated that a magnetic field vector that is aligned with the observer's line-of-sight ($\gamma=0,180\deg$) can be easily
confused with a more perpendicular one ($\gamma \approx 90\deg$) owing to the effect of the photon noise. In that experiment, the signals in the
linear polarization were solely due to the photon noise. We did not study whether the inversion code could retrieve useful
information about $\gamma$ from the Stokes $Q$ and $U$ profiles that, in spite of being below the $4.5\sigma$-level, introduce a correlation within 
the noise. To study this possibility, we carry out a similar experiment to the one we have performed in Section~\ref{subsection:phinoise}, but 
focusing instead on the inclination of the magnetic field vector with respect to the observer's line-of-sight $\gamma$. To this end, we prescribe 
a theoretical distribution of the magnetic field vector in the form\\

\begin{eqnarray}
\mathcal{P}_1(\ve{B},\alpha) \df\ve{B} \df\alpha = \frac{1}{2 \pi B_0} \delta\left(\gamma-\frac{\pi}{4}\right) \textrm{H}(B-B_0) \df B \df\gamma \df\phi 
\df\alpha \;,
\label{equation:pdfdeltagamma}
\end{eqnarray}

\noindent which is identical to the distribution in Equation~\ref{equation:pdfdeltaazim} but where the roles of $\phi$ and $\gamma$
are exchanged. That is, the distribution of the azimuth is now uniform and the distribution of the inclination is now a $\delta$-Dirac function
centered at $\gamma = 45\deg = \pi/4$ rad, which is chosen to ensure that $B_\parallel= B \cos\gamma$ and $B_\perp= B \sin\gamma$ 
(see Eq.~\ref{equation:wfa}) are equal. As we did in Section~\ref{subsection:gammanoise}, we add to the resulting Stokes profiles synthesized with
Equation~\ref{equation:pdfdeltagamma} photon noise with the values $\sigma=10^{-3}, 3\times 10^{-4}$. We also apply the three
selection criteria mentioned above of $S/R_{\rm quv}$, $S/R_{\rm qu}$, and $S/R_{\rm no-qu}$. The inferred histograms of $\gamma$ are
displayed in Figure~\ref{figure:gamma}.\\

With the first selection criterion, $S/R_{\rm quv}$, we realize that the inversion code retrieves a distribution of $\gamma$ (solid lines in 
Fig.~\ref{figure:gamma}) with a clear peak at the original value of $\gamma=45\deg$. This distribution has however an extended asymmetric 
tail towards higher values of $\gamma$. This indicates that there is a tendency to retrieve field that are more inclined than they originally are, as a
consequence of inverting Stokes profiles where $Q$ and $U$ are not 4.5 times above the noise level. This becomes clear when we consider the 
$S/R_{\rm qu}$-criterion (dashed lines in Fig.~\ref{figure:gamma}), which ensures that the linear polarization profiles are above the 
$4.5\sigma$-level, since in this case the tail towards larger values of $\gamma$ disappears. What are the properties of the peak at $\gamma=45\deg$ obtained 
with the $S/R_{\rm quv}$-criterion ? Is it a consequence of the inversion code being able to properly retrieve the inclination 
from the weak linear-polarization signals that are below the noise
level but introduce a correlation into the noise ? Or does the peak contain, as in the case of $\phi$ (Sect.~\ref{subsection:phinoise})
and since a large portion of the Stokes profiles were selected with the $S/R_{\rm quv}$-criterion, strong linear-polarization signals ? This can be answered
by the $S/R_{\rm no-qu}$-criterion, which only selects Stokes profiles for which Stokes $Q$ and $U$ are below the $4.5\sigma$-level (see dotted-dashed
lines in Fig.~\ref{figure:gamma}). This selection criteria shows an even more pronounced tail towards larger values of $\gamma$ than the $S/R_{\rm quv}$-criteria.
This is a consequence of many more Stokes profiles containing no information about $\gamma$ owing to very noisy linear-polarization signals. However,
the peak at $\gamma=45\deg$ does not completely disappear, which indicates that, unlike the case of $\phi$, the inversion code is able to partially use the 
correlation hidden in the noise as a means of determining $\gamma$.\\

We note, however, that this does not mean that the inversion code is able to retrieve the correct distribution of $\gamma$ by employing
the $S/R_{\rm no-qu}$-criterion (Stokes $Q$ and $U$ profiles below the $4.5\sigma$-level) when the distribution is a more general 
one than the $\delta$-Dirac function employed in Equation~\ref{equation:pdfdeltagamma}. To emphasize this statement, it suffices 
to point out that not even the $S/R_{\rm quv}$-criteria, where 20-40 \% of the Stokes profiles have linear polarization signals above
the $4.5\sigma$-level, is able to do so (see Figs.~\ref{figure:highquv}d-\ref{figure:lowquv}d). As demonstrated in Section~\ref{section:discussion},
the only way of retrieving the original distributions of the magnetic field vector is to ensure that the linear polarization is above the
$4.5\sigma$-level, that is, to apply the $S/R_{\rm qu}$-criterion.\\

\section{Conclusions}
\label{section:conclu}

We have carried out several numerical experiments in which a large number of Stokes profiles have been synthesized using
uniform distributions of the three components of the magnetic field vector $B$, $\gamma$, and $\phi$, and the magnetic filling 
factor $\alpha$ (Equation~\ref{equation:pdfuniform}). To these Stokes profiles, we have added photon noise to two different levels of
$\sigma=10^{-3}$, and $\sigma=3\times 10^{-4}$. We have then inverted these profiles in order to retrieve the original 
distributions employed in the synthesis. This has been done in two different ways. In the first one, we have selected the pixels 
where the $S/R$ ratio in any of the polarization signals (circular or linear) are equal to or larger than 4.5, the so-calld $S/R_{\rm quv}$-criterion. In the 
second case, we have selected the pixels where the $S/R$ ratio in the linear polarization signals is equal to or larger 
than 4.5, which is the $S/R_{\rm qu}$-criterion. The former criterion, $S/R_{\rm quv}$, selects Stokes profiles that arise from all possible
ranges of $B$, $\gamma$, and $\phi$ (dashed-black curves in Fig.~\ref{figure:highquv}-\ref{figure:lowquv}), whereas the $S/R_{\rm qu}$-criterion 
selects Stokes profiles that arise from highly inclined magnetic fields (dashed-black curves in Fig.~\ref{figure:highqu}-\ref{figure:lowqu}).\\ 

In addition, we have demonstrated (see Section~\ref{section:discussion}) that the $S/R_{\rm qu}$-criterion
can recover, with much larger reliability than $S/R_{\rm quv}$, the original distributions of the three components of the magnetic
field vector from the selected profiles. In particular, the latter criterion clearly overestimates the inclination (with respect
to the observer's line-of-sight) of the  magnetic field vector $\gamma$ (solid-red lines in Figs.~\ref{figure:highquv}d-\ref{figure:lowquv}d), and 
therefore also overestimates the component of the magnetic field that is perpendicular to the observer's line-of-sight, $B_\perp$
(solid-red lines in Figs.~\ref{figure:highquv}b-\ref{figure:lowquv}b). To avoid these systematic errors, we propose employing instead the $S/R_{\rm qu}$-criterion, 
which allows the correct distribution of $B_\perp$ and $\gamma$ (solid-red lines in Figs.~\ref{figure:highqu}b,d-\ref{figure:lowqu}b,d) 
from the selected profiles to be retrieved. Unfortunately, as mentioned above, the $S/R_{\rm qu}$-criterion systematically selects the Stokes profiles
where the magnetic field vector is already rather inclined, and thus ends up having the same sort of bias as the $S/R_{\rm quv}$-criterion. 
It seems therefore that the only way around this problem is to decrease the photon noise to a sufficiently low level that the vast majority
of profiles have a $S/R$ ratio larger than 4.5 in the linear polarization profiles. In this way, the selected Stokes profiles
would arise from magnetic fields that are representative of the real distribution. It is important to bear in mind that the overestimation 
of $\gamma$ and $B_\perp$ is a result that does not depend on the probability distribution function employed in our tests (Eq.~\ref{equation:pdfuniform}).\\

We have also studied the ability of the inversion code to retrieve the angular distribution of the magnetic field
vector ($\gamma$ and $\phi$), even when the Stokes $Q$ and $U$ profiles are below the $4.5\sigma$-level (Section~\ref{section:correlations}).
In this case, we employed initial distributions of $\phi$ and $\gamma$ featuring $\delta$-Dirac functions. We found that the 
correlation introduced into the noise by these signals does not help us to retrieve the correct azimuthal angle $\phi$. Instead of the original
$\delta$-function, the inversion code yields a random distribution of azimuthal angles that are uniformly distributed (Fig.~\ref{figure:azimuth}). 
In the case of the inclination of the magnetic field vector $\gamma$, the inversion code is able to partially retrieve the $\delta$-function from
the weak signals below the noise (Fig.~\ref{figure:gamma}). However, this is not the case when employing more general distributions, in which case
the aforementioned systematic overestimation of $\gamma$, dominates the inferred distribution (cf. del Toro Iniesta et al. 2010).\\

We caution that in our experiments we employed only one spectral line, Fe I ($g_{\rm eff}=2.5$) 6302.5 {\AA}, whereas in many 
previous studies Fe I ($g_{\rm eff}=1.67$) 6301.5 {\AA} had also been analyzed. Employing two spectral lines would certainly somewhat improve 
the results from these experiments as there would be more data points available to the inversion code. However, based 
on the results of Orozco Su\'arez et al. (2010), these improvements will likely be of second order compared to the systematic errors 
presented in this paper.\\

An additional point of consideration refers to the numerical code used to infer the magnetic field vector from the inversion of the radiative
transfer equation for polarized light (VFISV). This code operates, with some modifications, under a Levenberg-Marquardt non-linear minimization
algorithm (Borrero et al. 2010). We cannot exclude that other numerical schemes such as genetic (Lagg et al. 2004) or Bayesian
algorithms (Asensio Ramos et al. 2008; Asensio Ramos 2009) are less affected by photon noise and therefore able to infer more accurately the
distribution of the magnetic field vector.\\

\begin{acknowledgements}
This research has greatly benefited from discussions that were held at the International Space Science 
Institute (ISSI) in Bern (Switzerland) in February 2010 as part of the International Working group \emph{Extracting
Information from spectropolarimetric observations: comparison of inversion codes}. This work has made use
of the NASA Astrophysical Data System.
\end{acknowledgements}

\end{document}